\documentclass[twocolumn,preprintnumbers,amsmath,amssymb,nofootinbib,longbibliography]{revtex4-1}

\usepackage{graphicx}
\usepackage{dcolumn}
\usepackage{bm}
\usepackage{color}
\usepackage{lipsum}

\usepackage{ulem}

\usepackage{hhline}

\usepackage[a4paper]{hyperref}
\hypersetup{colorlinks=true,linktoc=all,linkcolor=blue,breaklinks=true,citecolor=blue,urlcolor=blue}

\newcommand{\zub}[2]{\langle\!\langle #1\vert #2 \rangle\!\rangle}

\newcommand{\be}{\begin{equation}}
\newcommand{\ee}{\end{equation}}
\newcommand{\bea}{\begin{eqnarray}}
\newcommand{\eea}{\end{eqnarray}}

\newcommand{\eq}[1]{Eq.~(\ref{#1})}
\newcommand{\fig}[1]{Fig.~\ref{#1}}
\newcommand{\figs}[1]{Figs.~\ref{#1}}

\newcommand{\e}{\varepsilon}

\newcommand{\new}[1] {{{#1}}}

\begin{document}

\title{Splitting efficiency and interference effects in a Cooper pair splitter \\
based on a triple quantum dot with ferromagnetic contacts}

\author{Kacper Bocian}
\email{bocian@amu.edu.pl}
\affiliation{Faculty of Physics, Adam Mickiewicz University, 61-614 Pozna\'n, Poland}

\author{Wojciech Rudzi\'nski}
\affiliation{Faculty of Physics, Adam Mickiewicz University, 61-614 Pozna\'n, Poland}

\author{Ireneusz Weymann}
\affiliation{Faculty of Physics, Adam Mickiewicz University, 61-614 Pozna\'n, Poland}

\date{\today}

\begin{abstract}
We theoretically study the spin-resolved subgap transport properties of a Cooper pair splitter based on a triple quantum dot attached to superconducting and
ferromagnetic leads. Using the Keldysh Green's function formalism, we analyze the dependence of the Andreev conductance, Cooper pair splitting efficiency, and tunnel
magnetoresistance on the gate and bias voltages applied to the system. We show that the system's transport properties are strongly affected by spin dependence of
tunneling processes and quantum interference between different local and nonlocal Andreev reflections. We also study the effects of finite hopping between the side
quantum dots on the Andreev current. This allows for identifying the optimal conditions for enhancing the Cooper pair splitting efficiency of the device. We find
that the splitting efficiency exhibits a nonmonotonic dependence on the degree of spin polarization of the leads and the magnitude and type of hopping between the
dots. An almost perfect splitting efficiency is predicted in the nonlinear response regime when the energies of the side quantum dots are tuned to the energies of
the corresponding Andreev bound states.
\new{
In addition, we analyzed features of the tunnel magnetoresistance (TMR) for
a wide range of the gate and bias voltages, as well as for different model parameters,
finding the corresponding sign changes of the TMR in certain transport regimes.
}%
The mechanisms leading to these effects are thoroughly discussed.
\end{abstract}

\maketitle

%%%%%%%%%%%%%%%%%%%%%%%%%%%%%%%%%%%%%%%%%%%%%%%%%%%%%%
%%%%%%%%%%%%%%%%%%%%%%%%%%%%%%%%%%%%%%%%%%%%%%%%%%%%%%
\section{Introduction}
%%%%%%%%%%%%%%%%%%%%%%%%%%%%%%%%%%%%%%%%%%%%%%%%%%%%%%
%%%%%%%%%%%%%%%%%%%%%%%%%%%%%%%%%%%%%%%%%%%%%%%%%%%%%%

Hybrid nanostructures, involving normal and superconducting parts, have recently become a subject of extensive studies in the context of generation and manipulation of
pairs of quantum-entangled objects \cite{Loss1998Jan,Ruggiero2006,Ciudad2015May}. In particular, efforts stimulated by development of solid-state quantum information
devices have given rise to various realizations of Cooper pair splitters (CPS)
\cite{Hofstetter2009Oct,Hofstetter_PhysRevLett.107.136801,Herrmann_PhysRevLett.104.026801,Das2012Nov,Schindele_PhysRevLett.109.157002,Fulop_PhysRevB.90.235412,Tan_PhysRevLett.114.096602,
Fulop_PhysRevLett.115.227003}. Most of CPS setups consist of a superconductor, which serves as a source of pairs of entangled electrons, coupled to two normal metal
drain contacts by means of tunable quantum dots. The advantage of this configuration is that the process of Cooper pair splitting can be controlled by appropriate gate
voltages attached to the dots. For sufficiently low temperatures and voltages smaller than the superconducting energy gap, transport through the system occurs mainly
through Andreev reflection processes \cite{andreev,Recher2001Apr,DeFranceschi2010,MartinRodero_AdinPhys2011}. Cooper pair electrons can be transferred either through
one arm of the device in a direct Andreev reflection (DAR) process or through two arms of the splitter in a crossed Andreev reflection (CAR) process. The latter
processes are in fact the ones that make the device work as a Cooper pair beam splitter \cite{Hofstetter2009Oct}. Therefore, it is important to optimize the device
parameters in such a manner that CAR processes are maximized \cite{Das2012Nov,Schindele_PhysRevLett.109.157002}.

The rate of DAR and CAR processes strongly depends on both the on-dot and inter-dot Coulomb correlations \cite{Hofstetter2009Oct}. In particular, in the case of
considered quantum-dot-based splitters, it is important to have the on-dot correlations much larger than the intradot ones. Then, CAR processes dominate Andreev
transport and the device is characterized by a large Cooper pair splitting efficiency $\eta$. The splitting efficiency also greatly depends on the transport regime and
the position of the levels of the quantum dots. More specifically, the Andreev reflection processes become generally reduced when the system is detuned from the
particle-hole symmetry point. In addition, it is also possible to affect the magnitude of Andreev conductance by changing the ratio of the couplings to normal leads
and to the superconductor \cite{Zhu_PRB01,Dominguez2016,Hussein2017,Hussein2016,Michalek2017}. All this clearly demonstrates that there are many tunable parameters,
which allow for optimizing the operation of a CPS device.

The transport properties of quantum-dot-based CPS are already relatively well understood \cite{Loss2000,Recher2001Apr, Saraga2003,
PhysRevB.81.094526,Tanaka2010Feb,Eldridge2010Nov, Chevallier2011_PhysRevB.83.125421, PhysRevB.85.035419,Soller2012,Trocha2014_PhysRevB.89.245418,
Sothmann2014Dec,Trocha2015Jun,Burset2011,Amitai2016,Gong2016,WrzesniewskiJPCM17}. Such systems can be modeled by a double quantum dot Anderson-type Hamiltonian, with the two dots
coupled to a common superconducting lead and each dot attached to a normal electrode. However, recent experiments of F\"ul\"op {\it at al.}
\cite{Fulop_PhysRevLett.115.227003} have shown that such modeling may be insufficient to describe certain subtle effects resulting from the quantum interference
between different Andreev reflection tunneling events. To properly account for such effects, it has been suggested \cite{Fulop_PhysRevLett.115.227003} that one needs
to resort to a three-site model, in which there are two quantum dots in the arms of the splitter, while a large, middle quantum dot is formed in a direct proximity of
the superconductor.

The transport properties of quantum-dot-based splitters modeled by appropriate triple quantum dot Hamiltonian have been in fact recently considered in the case of
three dots coupled to one superconducting and two normal, nonmagnetic electrodes \cite{Dominguez2016}. In this paper, we extend these studies and analyze the
quantum interference effects and the splitting efficiency of CPS based on triple quantum dots attached to ferromagnetic contacts, the problem which still remains
rather unexplored. Aside from the possibility to tune the ratio of DAR and CAR processes by spin polarization of the leads
\cite{Trocha2014_PhysRevB.89.245418,Trocha2015Jun}, ferromagnetic electrodes were shown to be crucial in detection of entanglement between split Cooper pair electrons
\cite{Klobus2014Mar,Busz2017Aug}. Our studies are performed by using the Keldysh Green's function approach, which allows for analyzing the impact of interference
effects on Andreev conductance and splitting efficiency in both the linear and nonlinear response regimes. In addition, we examine the impact of direct hopping between
the dots forming the arms of the splitter on the spin-resolved transport properties of the system. Furthermore, the effects of the Rashba spin-orbit interaction
\cite{Sun2005,Droste_JPCM12,Bai2012,Nian2014,Hussein2016,Shekhter2016,Hussein2017} on the splitting efficiency $\eta$ of the device are also discussed. We show that
$\eta$ greatly depends on the arrangement of magnetic moments of ferromagnetic leads and the positions of the quantum dots' levels. We also demonstrate a rather
detrimental effect of the spin-orbit coupling on the Cooper pair splitting efficiency.
\new{
Finally, we predict that the dependence of $\eta$
on the degree of spin polarization of the leads can be strongly
modified by finite amplitude of hopping between the quantum dots located in the arms of the splitter.}

The paper is organized as follows. Section~\ref{Sec:II} presents the theoretical framework of the paper, where the Hamiltonian (Sec.~\ref{Sec:IIA}), method
(Sec.~\ref{Sec:IIB}), and quantities of interest (Sec.~\ref{Sec:IIC}) are described. Results in the linear response regime are presented and discussed in
Sec.~\ref{Sec:III}. First, the Andreev conductance is analyzed (Sec.~\ref{Sec:IIIA}), then the tunnel magnetoresistance (TMR) is studied (Sec.~\ref{Sec:IIIB}),
followed by splitting efficiency (Sec.~\ref{Sec:IIIC}). The nonlinear response regime is analyzed in Sec.~\ref{Sec:IV}, with Secs.~\ref{Sec:IVA}, \ref{Sec:IVB}, and
\ref{Sec:IVC} devoted to Andreev conductance, TMR and splitting efficiency, respectively. The summary and conclusions can be found in Sec.~\ref{Sec:V}.

%%%%%%%%%%%%%%%%%%%%%%%%%%%%%%%%%%%%%%%%%%%%%%%%%%%%%%
\section{Theoretical formulation}
\label{Sec:II}
%%%%%%%%%%%%%%%%%%%%%%%%%%%%%%%%%%%%%%%%%%%%%%%%%%%%%%
\subsection{Model and Hamiltonian}
\label{Sec:IIA}
%%%%%%%%%%%%%%%%%%%%%%%%%%%%%%%%%%%%%%%%%%%%%%%%%%%%%%

\begin{figure}[t]
    \includegraphics[width=1\columnwidth]{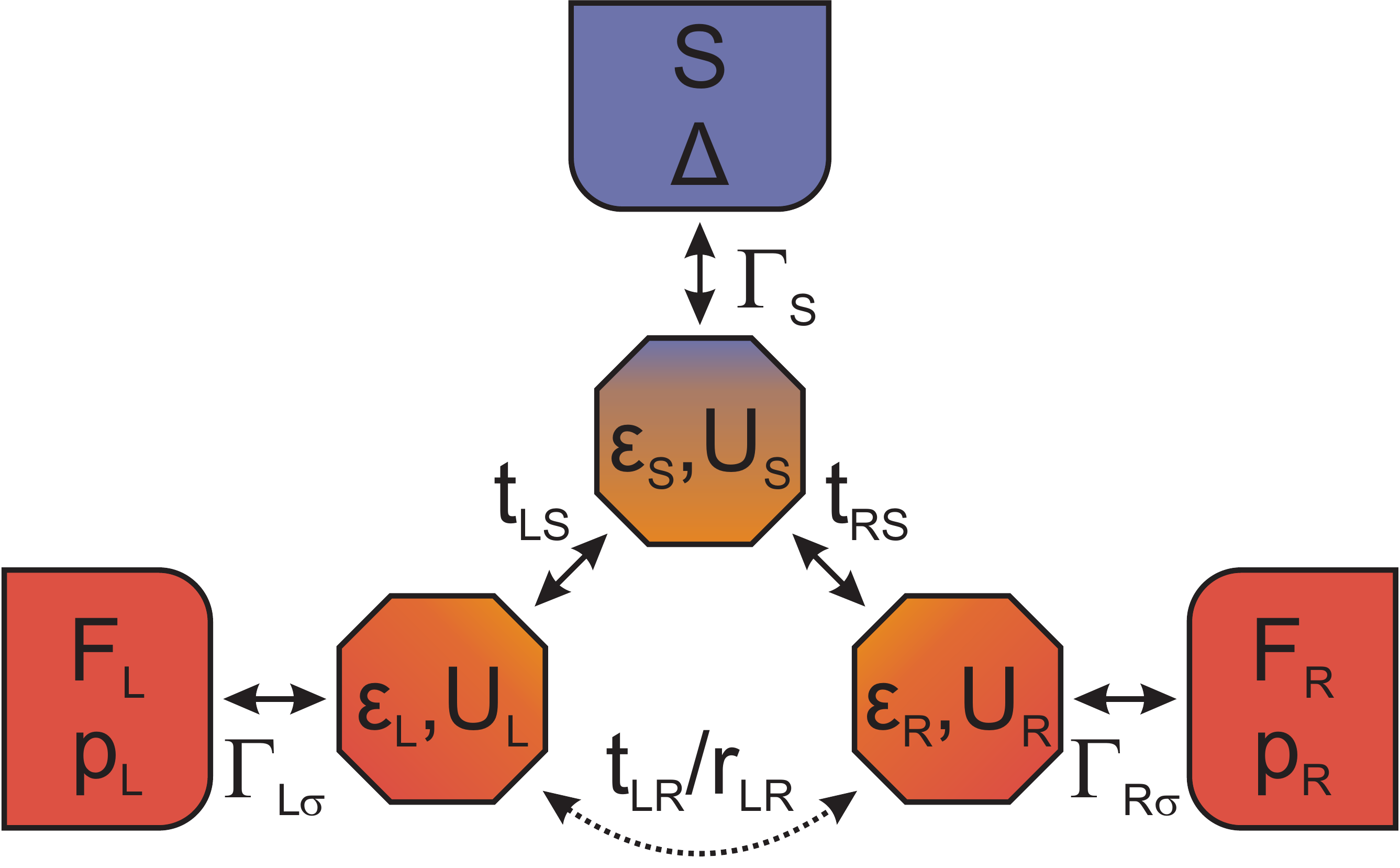}
    \caption{
        The schematic representation of the quantum-dot-based Cooper pair splitter
        with ferromagnetic contacts.
        The system consists of three quantum dots, where the middle dot ($j=S$) is attached
        to superconducting (S) electrode, while the left ($j=L$) and right ($j=R$) dot is coupled to
        the corresponding ferromagnetic (F) lead.
        For details, see the main text.
    }
    \label{Fig:model}
\end{figure}

Our investigations are focused on a triple quantum dot based
Cooper pair splitter with ferromagnetic electrodes,
as schematically shown in Fig.~\ref{Fig:model}.
Each quantum dot is coupled to a separate electrode,
with the middle dot attached to an $s$-wave superconductor
and the left (right) dot coupled to the corresponding left (right)
ferromagnetic electrode. The magnetic moments of the ferromagnets
are assumed to form either parallel or antiparallel magnetic configuration.
The two side dots together with the leads
form thus the arms of the Cooper pair splitter.
The entire system can be described by the
following Hamiltonian \cite{Fulop_PhysRevLett.115.227003,Dominguez2016}
\begin{equation}\label{eq:ham}
    H=H_{\rm F}+H_{\rm S}+H_{\rm QDs}+H_{\rm T}.
\end{equation}
The first term $H_{\rm F}$ describes the noninteracting electrons
in the left ($j=L$) and right ($j=R$) ferromagnetic lead,
\begin{equation}
    H_{\rm F}=\sum_{j=L,R}\sum_{k\sigma}\varepsilon_{j k\sigma}f_{j k\sigma}^{\dagger}f_{j k\sigma},
\end{equation}
where $f_{j k\sigma}^{\dagger}$
($f_{j k\sigma}$) stands for the creation (annihilation) operator
of an electron in the $j$-th lead with the wave vector $k$,
spin $\sigma$ and the energy $\varepsilon_{j k\sigma}$.
The next term of the total Hamiltonian, $H_{\rm S}$,
describes the $s$-wave superconducting lead modeled by
\begin{equation}
    H_{\rm S}=\sum_{p\sigma}\varepsilon_{p} s_{p\sigma}^{\dagger}s_{p\sigma}
    +\Delta\sum_{p}\left(s_{-p\uparrow}^{\dagger}s_{p\downarrow}^{\dagger}+H.c.\right),
\end{equation}
with $s_{p\sigma}^{\dagger}$ ($s_{p\sigma}$) creating (annihilating) an electron
with the wave vector $p$, spin $\sigma$ and energy $\varepsilon_{p}$.
The second part of $H_{\rm S}$ characterizes the superconducting energy gap $\Delta$.
The three quantum dots are described by
\begin{eqnarray}
    H_{\rm QDs} &=& \sum_{j=L,R,S}\left(\varepsilon_{j}n_j
    +U_{j}n_{j\uparrow} n_{j\downarrow}\right) \nonumber\\
    & + & \sum_{\sigma}\big(t_{LR} d_{L\sigma}^{\dagger}d_{R\sigma} + \sum_{j=L,R} t_{jS} d_{j\sigma}^{\dagger}d_{S\sigma} + H.c.\big)\nonumber\\
    & - & \sum_{\sigma}\big(R_{LR}d_{L\bar{\sigma}}^{\dagger}\left(i\boldsymbol{\sigma}_{x}\right)_{\sigma\bar{\sigma}}d_{R\sigma} + H.c.\big),
\end{eqnarray}
where $n_j = n_{j\uparrow}+n_{j\downarrow}$, with $n_{j\sigma} = d_{j\sigma}^{\dagger}d_{j\sigma}$ and $d_{j\sigma}^{\dagger}$ ($d_{j\sigma}$) being the creation
(annihilation) operator of an electron with spin $\sigma$ and energy $\varepsilon_{j}$ in the $j$-th dot. The on-site Coulomb correlations on the $j$-th dot are
described by $U_{j}$. The second line accounts for the hopping between the dots, with $t_{ij}$ indicating the corresponding hopping amplitude. The last term describes
the Rashba spin-orbit type of coupling, which can be understood as a spin-flip hopping between the left and right dots with amplitude $R_{LR}$. The symbol
$\boldsymbol{\sigma}_{x}$ is the Pauli spin matrix along the $x$ axis. Here, the quantization axis is assumed to coincide with the magnetic moments of the
ferromagnetic electrodes, which are assumed to be collinear to each other. It is worth noting that we assume that the hopping amplitudes are real and symmetric,
$t_{ij}=t_{ji}$ and $R_{LR}=R_{RL}$. For the further study we introduce the complex spin-flip amplitudes, which satisfy the condition $r_{LR}=-iR_{LR}$, such that
$r_{RL}=-r_{LR}$.

Finally, the last term of the total Hamiltonian, $H_{\rm T}$,
describes tunneling processes between the corresponding leads and quantum dots
\begin{eqnarray}
    H_{\rm T} \! = \!\!\! \sum_{j=L,R} \sum_{k\sigma} T_{k\sigma}^{j} f_{j k\sigma}^{\dagger}d_{j\sigma}
    + \sum_{p\sigma}T_{p}^{S}s_{p\sigma}^{\dagger}d_{S\sigma}+H.c.
\end{eqnarray}
Here, $T_{k\sigma}^{j}$ ($T_{p}^{S}$) denotes the tunneling amplitude between the $j$-th quantum dot and the corresponding ferromagnetic (superconducting) lead. The
amplitudes result in the broadening of the energy levels of the dots, which is described by, $\Gamma_{j\sigma} = 2\pi \rho_{j\sigma} | T_{k\sigma}^{j} |^{2}$, for
ferromagnetic leads and, $\Gamma_{S} = 2\pi\rho_{S} | T_{p}^{S}|^{2}$, for superconducting lead, where $\rho_{j\sigma}$ and $\rho_{S}$ are the densities of states of
the corresponding leads taken at the Fermi energy. For ferromagnetic leads, one can express the couplings in terms of spin polarization of given lead $p_j$ as,
$\Gamma_{j\sigma} = \Gamma_{j} (1+\hat{\sigma} p_j )$, where $\hat{\sigma}=\pm 1$ for $\sigma=\uparrow,\downarrow$.
\new{
The spin polarization is defined as, $p_j = (\rho_{j\uparrow} - \rho_{j\downarrow}) / (\rho_{j\uparrow} + \rho_{j\downarrow})$, where the spin up in the parallel
configuration is considered as belonging to the majority-spin subband of a given lead. In the antiparallel magnetic configuration of the device, the magnetization of
the right dot is flipped and the spin-up (spin down) electrons correspond to the minority- (majority-) spin subband. }

%%%%%%%%%%%%%%%%%%%%%%%%%%%%%%%%%%%
\subsection{Method}
\label{Sec:IIB}
%%%%%%%%%%%%%%%%%%%%%%%%%%%%%%%%%%%

To calculate the transport characteristics, we employ the nonequilibrium Green's function technique in the $12\times 12$ Nambu representation. Accordingly, by using
the equation of motion method, we derived the expressions for the Fourier transform of the retarded Green's function $\boldsymbol{G}^{r}\left(\omega\right)=\zub
{\boldsymbol{\Psi}} { \boldsymbol{\Psi}^{\dagger}}$, where the Nambu spinor takes the form $\boldsymbol{\Psi}=\big(d_{L\uparrow}^{\dagger}, d_{L\downarrow},
d_{S\uparrow}^{\dagger}, d_{S\downarrow}, d_{R\uparrow}^{\dagger}, d_{R\downarrow}, d_{L\downarrow}^{\dagger}, d_{L\uparrow}, d_{S\downarrow}^{\dagger}, d_{S\uparrow},
d_{R\downarrow}^{\dagger}, d_{R\uparrow}\big)^{\!\dagger}$.
For example, the equation of motion for $\zub{d_{L\sigma}} {d_{j\sigma'}^{\dagger}}$ (where $j=L,R,S$ and $\sigma'=\uparrow,\downarrow$) reads as
\begin{eqnarray} \label{eq:MainEOM}
    & &\left(\omega-\varepsilon_{L}\right) \zub{ d_{L\sigma} }{ d_{j\sigma'}^{\dagger}} = \delta_{Lj}\delta_{\sigma\sigma'} +
    U_{L} \zub {d_{L\sigma}d_{L\bar{\sigma}}^{\dagger}d_{L\bar{\sigma}}} { d_{j\sigma'}^{\dagger} } + \nonumber\\
    & &+\; \sum_{j'=R,S} t_{Lj'} \zub{ d_{j'\sigma} }{ d_{j\sigma'}^{\dagger} } + r_{LR} \zub{ d_{R\bar{\sigma}} }{ d_{j\sigma'}^{\dagger} } + \nonumber\\
    & &+\; \sum_{k}T_{k\sigma}^{L*} \zub{ f_{Lk\sigma} }{ d_{j\sigma'}^{\dagger} }.
\end{eqnarray}
Equation~(\ref{eq:MainEOM}) involves new correlation functions, including the Green's functions of higher order, for which one needs to write a new set of equations of
motion, accordingly:
\begin{eqnarray} \label{eq:hoGF}
    & & \left(\omega-\varepsilon_{L}-U_L\right) \zub {d_{L\sigma}d_{L\bar{\sigma}}^{\dagger}d_{L\bar{\sigma}} } { d_{j\sigma'}^{\dagger} } =\nonumber\\
    & & \delta_{Lj} \delta_{\sigma\sigma'} \langle d_{L\bar{\sigma}}^{\dagger}d_{L\bar{\sigma}} \rangle
    - \delta_{Lj} \delta_{\bar{\sigma}\sigma'} \langle d_{L\bar{\sigma}}^{\dagger}d_{L\sigma} \rangle +\nonumber\\
    & & +\sum_{j'=R,S} \left[ t_{Lj'} \zub {d_{L\bar{\sigma}}^{\dagger}d_{L\bar{\sigma}}d_{j'\sigma} }{ d_{j\sigma'}^{\dagger} } \right. +\nonumber\\
    & & +t_{Lj'} \zub{ d_{L\sigma}d_{L\bar{\sigma}}^{\dagger}d_{j'\bar{\sigma}} }{ d_{j\sigma'}^{\dagger} } +\nonumber\\
    & & +\left. t_{Lj'} \zub{ d_{L\sigma}d_{L\bar{\sigma}}d_{j'\bar{\sigma}}^{\dagger} }{ d_{j\sigma'}^{\dagger} } \right] +\nonumber\\
    & & +r_{LR} \zub{ d_{L\bar{\sigma}}^{\dagger}d_{L\bar{\sigma}}d_{R\bar{\sigma}} }{ d_{j\sigma'}^{\dagger} } +\nonumber\\
    & & +r_{LR} \zub{ d_{L\sigma}d_{L\bar{\sigma}}^{\dagger}d_{R\sigma} }{ d_{j\sigma'}^{\dagger} } +\nonumber\\
    & & -r_{LR} \zub{ d_{L\sigma}d_{L\bar{\sigma}}d_{R\sigma}^{\dagger} }{ d_{j\sigma'}^{\dagger} } +\nonumber\\
    & & +T_{k\sigma}^{L*} \zub{ f_{Lk\sigma}d_{L\bar{\sigma}}^{\dagger}d_{L\bar{\sigma}} }{ d_{j\sigma'}^{\dagger} } +\nonumber\\
    & & +T_{k\bar{\sigma}}^{L} \zub{ f_{Lk\bar{\sigma}}^{\dagger}d_{L\sigma}d_{L\bar{\sigma}} }{ d_{j\sigma'}^{\dagger} } +\nonumber\\
    & & +T_{k\bar{\sigma}}^{L*} \zub{ f_{Lk\bar{\sigma}}d_{L\sigma}d_{L\bar{\sigma}}^{\dagger} }{ d_{j\sigma'}^{\dagger} },
\end{eqnarray}
and
\begin{eqnarray} \label{eq:hoGF2}
    \left(\omega-\varepsilon_{Lk\sigma}\right) \zub{ f_{Lk\sigma} }{ d_{j\sigma'}^{\dagger} } =
    T_{k\sigma}^{L} \zub{ d_{L\sigma} }{ d_{j\sigma'}^{\dagger} },
\end{eqnarray}
where $\langle \ldots \rangle$ denotes the expectation value.

By writing the next equations of motion for the new
Green's functions appearing in Eq.~(\ref{eq:hoGF}),
one again generates new Green's functions of even higher order.
Therefore, to close the set of equations, at this stage we make
use of the so-called Hubbard-I approximation \cite{Pals1996,PhysRevB.88.155425,Fulop_PhysRevLett.115.227003},
which simplifies the higher-order Green's functions.
\new{More specifically,
we apply the following general form of the Green's function decoupling scheme, $\langle\langle\boldsymbol{ABC}|\boldsymbol{D}\rangle\rangle=
\langle\boldsymbol{AB}\rangle\langle\langle\boldsymbol{C}|\boldsymbol{D}\rangle\rangle+
\langle\boldsymbol{BC}\rangle\langle\langle\boldsymbol{A}|\boldsymbol{D}\rangle\rangle-
\langle\boldsymbol{AC}\rangle\langle\langle\boldsymbol{B}|\boldsymbol{D}\rangle\rangle$, with $\boldsymbol{A}$, $\boldsymbol{B}$, $\boldsymbol{C}$, and
$\boldsymbol{D}$ denoting the corresponding fermion operators in the model Hamiltonian considered here. The latter gives rise to the following approximations for the
formula (\ref{eq:hoGF}).
First, we assume that the couplings between the side dots and the external ferromagnetic leads are relatively weak, which implies vanishing of the following
expectation values: $\langle f_{Lk\sigma}^{(\dag)}d_{L\sigma(\bar{\sigma})}^{(\dag)}\rangle\approx 0$. Second, we assume that the on-dot and interdot spin relaxation
processes are negligible, such that $\langle d_{L\sigma}^{\dagger}d_{L(R)\bar{\sigma}}\rangle\approx0$. Third, since in out setup only the middle dot is proximitized
by the superconductor, we consider the expectation values, $\langle d_{L\sigma}d_{j\sigma(\bar{\sigma})}\rangle$, with $j=R,S$, to be negligibly small.
Using similar approximations, we set up
the equations of motion for the other correlators,
which allows us to close the set of equations and to obtain the full Green's function
$\boldsymbol{G}^{r}\left(\omega\right)$.}

The entire retarded Green's function can be written in the form of the Dyson's matrix equation
\begin{equation}
    \boldsymbol{G}^{r}\left(\omega\right)=\left\{ \left[\boldsymbol{g}^{r}
    \left(\omega\right)\right]^{-1}-\boldsymbol{\Sigma}^{r}\left(\omega\right)\right\} ^{-1},
\end{equation}
where $\boldsymbol{g}^{r}\left(\omega\right)$ is the Green's function
of the unperturbed system and $\boldsymbol{\Sigma}^{r}\left(\omega\right)$
denotes the self-energy matrix.
In order to study the system's transport properties,
one also needs to determine the lesser correlation function
$\boldsymbol{G}^{<}\left(\omega\right)$, which can be found
by using the Keldysh equation \cite{Keldysh1965,Jauho_book}
\begin{equation}
    \boldsymbol{G}^{<}\left(\omega\right)=
    \boldsymbol{G}^{r}\left(\omega\right)\boldsymbol{\Sigma}^{<}\left(\omega\right)\boldsymbol{G}^{a}\left(\omega\right),
\end{equation}
where $\boldsymbol{G}^{a}\left(\omega\right)=\left[\boldsymbol{G}^{r}\left(\omega\right)\right]^{\dagger}$ is the advanced Green's function and
$\boldsymbol{\Sigma}^{<}\left(\omega\right)$ stands for the lesser self-energy.
This self-energy can be approximated by the following equation
\begin{equation}
    \boldsymbol{\Sigma}^{<}\left(\omega\right)=i\sum_{j=L,R,S}\boldsymbol{f}_{j}\left(\omega\right)\boldsymbol{\Gamma}^{j}\left(\omega\right),
\end{equation}
where $\boldsymbol{f}_{j}\left(\omega\right)$ is the appropriate matrix of
the Fermi-Dirac distribution functions, while $\boldsymbol{\Gamma}^{j}\left(\omega\right)$
stands for the coupling matrix between the $j$-th quantum dot and the corresponding lead.

We note that the approximations made to decouple the Green's functions and close the set of equations for the determination of $\boldsymbol{G}^{r}\left(\omega\right)$
allow for resolving the effects of quantum interference on Andreev transport, such as the ones observed in Ref. \cite{Fulop_PhysRevLett.115.227003}, which are the main
focus of this paper. They are, however, insufficient to capture the strong electron correlations leading to the Kondo effect \cite{Goldhaber_Nature391/98}, and
especially the competition between Kondo and superconducting correlations \cite{Franke2011May,Domanski2016Mar,Wrzesniewski2017Nov}.

%%%%%%%%%%%%%%%%%%%%%%%%%%%%%%%%%%%
\subsection{Quantities of interest}
\label{Sec:IIC}
%%%%%%%%%%%%%%%%%%%%%%%%%%%%%%%%%%%

To calculate the current flowing from the $j$-th lead, one can use the Meir-Wingreen formula \cite{Meir_Phys.Rev.Lett.68/1992}
\begin{eqnarray*}
    I_{j} \!=\! \frac{ie}{h} \! \int \!\! d\omega\! \sum_{\sigma}\!
    \mathrm{Tr}\left[\boldsymbol{\Gamma}^{j}\! \left\{ \boldsymbol{f}_{j}(\omega)\!
    \left[\boldsymbol{G}^{r}(\omega)\!-\!\boldsymbol{G}^{a}
    (\omega)\right] \!+\! \boldsymbol{G}^{<} (\omega)\right\}\! \right].
\end{eqnarray*}
In the above equation, the trace indicates the summation over the electron component of the Nambu space.

\new{
Since in this paper we are mainly interested in the Andreev reflection processes
and the splitting properties of the device,
we assume that the superconducting energy gap is the largest energy scale in the problem.
In practice, we take the limit of infinite superconducting energy gap,
$\Delta\to \infty$, which allows us to focus exclusively
on the Andreev reflection processes \cite{Yoichi2007Jun,Domanski2016Mar}.}
Moreover, we assume the same bias voltage applied
to the ferromagnetic leads, $V_{L}=V_{R}=V$, while we
leave the superconducting lead floating, $V_{S}=0$.
In such a situation, the Andreev current flowing
through the ferromagnetic junction ($j=L,R$) can be explicitly found from
\cite{Fazio1998Mar,Sun1999Feb,Domanski2008Oct}
\begin{eqnarray}
    I_{j} & = & \frac{e}{h}\int d\omega\left[f\left(\omega-eV\right)-f\left(\omega+eV\right)\right] T_j(\omega),
\end{eqnarray}
where $f\left(\omega\right)=1/\left[\exp\left(\omega/T\right)+1\right]$, with the Boltzmann's constant $k_B\equiv 1$,
and $T_{j}(\omega)$ denotes the Andreev transmission coefficient
\begin{equation}
T_j(\omega) = \sum_\sigma \Big[ T_{j\sigma}^{\rm DAR}(\omega) + T_{j\sigma}^{\rm CAR}(\omega) \Big].
\end{equation}
Here, $T_{j\sigma}^{\rm DAR}(\omega)$ $[T_{j\sigma}^{\rm CAR}(\omega)]$
is the transmission coefficient due to direct (crossed) Andreev reflection processes.
These coefficients can be found from
\new{
\begin{equation} \label{eq:TDAR}
    T_{j\sigma}^{\rm DAR}(\omega) \!=\! \Gamma_{j\sigma}\! \left( \Gamma_{j\bar{\sigma}} |\zub{d_{j \sigma}}{d_{j \bar{\sigma}}}|^{2}
    \!+\! \Gamma_{j\sigma}  |\zub{d_{j \sigma}}{d_{j \sigma}}|^{2} \right) \!,
\end{equation}
\begin{equation} \label{eq:TCAR}
    T_{j\sigma}^{\rm CAR}(\omega) \!=\! \Gamma_{j\sigma}\! \left( \Gamma_{j'\bar{\sigma}} |\zub{d_{j \sigma}}{d_{j' \bar{\sigma}}}|^{2}
    \!+\! \Gamma_{j'\sigma} |\zub{d_{j \sigma}}{d_{j' \sigma}}|^{2}
\right),
\end{equation}}%
where if $j=L(R)$ then $j'=R(L)$. Note that the second parts of the above equations,
proportional to $\Gamma_{j\sigma}^2$ and $\Gamma_{j\sigma}\Gamma_{j'\sigma}$,
respectively, are non-zero only if the spin-orbit interaction is present.

The corresponding differential conductance is defined as, $G_{j}=dI_{j}/dV$,
which in the special case of the linear response regime may be written as
\begin{eqnarray} \label{eq:TotalG}
    G_{j} & = & \frac{2e^{2}}{h}\! \int \!\! d\omega\left(-\frac{\partial f\left(\omega\right)}{\partial\omega}\right) T_j(\omega).
\end{eqnarray}
The total current flowing between the ferromagnetic and superconducting leads can be found from, $I=I_{L}+I_{R}$, and consequently the total conductance is
$G=G_{L}+G_{R}$. Note also that with the formulas for Andreev transmission coefficient, we can inspect the separate contributions due to both DAR and CAR processes. We
can thus study the direct (crossed) Andreev current flowing through a given junction for a given spin, $I_{j\sigma}^{\rm DAR}$ ($I_{j\sigma}^{\rm CAR}$), together with
the total currents, $I^{\rm DAR} = \sum_{j=L,R}\sum_\sigma I_{j\sigma}^{\rm DAR}$ and $I^{\rm CAR} = \sum_{j=L,R}\sum_\sigma I_{j\sigma}^{\rm CAR}$, due to direct and
crossed Andreev reflections, respectively. In a similar fashion, one can analyze the corresponding contributions to the conductance, $G_{j\sigma}^{\rm DAR}$ and
$G_{j\sigma}^{\rm CAR}$, together with the total conductance due to DAR and CAR processes, $G^{\rm DAR} = \sum_{j=L,R}\sum_\sigma G_{j\sigma}^{\rm DAR}$ and $G^{\rm
CAR} = \sum_{j=L,R}\sum_\sigma G_{j\sigma}^{\rm CAR}$, respectively.

The ratio between the total currents flowing due to DAR and CAR processes
can be used to estimate the efficiency of the Cooper pair splitter, which can be defined as
\begin{equation}
    \eta = \frac{I^{\rm CAR}}{I^{\rm DAR} + I^{\rm CAR}} \times 100\%.
\end{equation}
If the current flows exclusively due to CAR processes, i.e., each Cooper pair leaving superconductor is split into two separate arms, the splitting efficiency is
maximum, $\eta = 100\%$. On the other hand, if only DAR processes drive the current, the splitting efficiency vanishes, $\eta = 0$.

In our considerations, we assume that the system is symmetric, $p_L = p_R \equiv p$, $U_L = U_R \equiv U$, $t_{jS}\equiv t_S$, $t_{LR} \equiv t$, $r_{LR} \equiv r$. We
also assume that the middle dot is much larger than the side dots, $U_S \ll \Gamma_S$, which implies that the middle dot can be treated as noninteracting. A similar
model and parameter set was in fact used to corroborate recent experimental results on transport through quantum-dot-based Cooper pair splitters with nonmagnetic
electrodes \cite{Fulop_PhysRevLett.115.227003}.

The goal of this paper is to analyze the role of spin-dependent tunneling
on the subgap transport behavior of Cooper pair splitters with ferromagnetic leads.
For that we consider two magnetic configurations of the device:
the parallel (P) configuration, in which magnetic moments of ferromagnets point in the same direction,
and the antiparallel (AP) configuration, in which the magnetization
of the right lead is flipped and the corresponding moments are opposite.
The system's magnetic configuration can be changed by applying a weak magnetic field
(much weaker than the critical field of the superconductor),
provided the two ferromagnets have different coercive fields. To estimate the change of transport
properties when the magnetic configuration is varied, we define the tunnel magnetoresistance
\cite{PhysRevB.79.054505,Weymann2014Mar,Trocha2015Jun}
\begin{equation} \label{eq:TMR}
    {\rm TMR} = \frac{I_{\rm AP} - I_{\rm P}} {I_{\rm P}} \times 100\%,
\end{equation}
where $I_{\rm P}$ ($I_{\rm AP}$) denotes the current flowing in the parallel (antiparallel) magnetic configuration. Note that because transferring a Cooper pair
involves two electrons of opposite spins, the Andreev current is usually larger in the antiparallel configuration compared to the parallel configuration, in which the
minority spin-channel is responsible for the reduction of tunneling, such that one typically finds $I_{\rm AP} > I_{\rm P}$
\cite{PhysRevB.79.054505,Weymann2014Mar,Trocha2015Jun}. Moreover, we would also like to notice that the rate of DAR processes does not depend on magnetic configuration
(since the two electrons of opposite spin tunnel to the same lead), while the rate of CAR processes does change when the configuration is varied. Consequently, the TMR
can provide additional information about the amount of CAR processes compared to DAR ones.
\new{Large values of the TMR can be considered as a signature
of an important role of CAR processes \cite{PhysRevB.79.054505,Weymann2014Mar,Trocha2015Jun}. One, however, needs to keep in mind that the TMR does not provide an
exact quantitative description of CAR and DAR processes and only together with other transport quantities sheds more light on the corresponding Andreev reflection
processes. In particular, a negative sign of TMR is mainly an indication of how the rate of CAR processes changes when the device's magnetic configuration is varied.}

For the considered system, we assume the Coulomb repulsion to be equal to $U=10\Gamma$,
whereas for the dot coupled to superconductor the Coulomb repulsions are neglected, $U_{S}=0$.
The couplings between the ferromagnetic leads and the corresponding
dots are symmetric, $\Gamma_{L}=\Gamma_{R}=\Gamma$,
while for the coupling between the middle dot and superconductor we assume $\Gamma_{S}=2\Gamma$.
Our results are calculated for the spin polarization of ferromagnets equal to $p=0.5$, if not stated otherwise.
The hopping between the middle and left (right) dot is assumed to be equal to $t_S=\Gamma/4$,
while all energies are measured in the unit of the coupling strength $\Gamma\equiv 1$.
The calculations are performed for the temperature $T=0.001\Gamma$.

\new{
We note that in the studied model only the middle dot is directly proximitized by the superconductor (see \fig{Fig:model}). Thus, one may expect that the Andreev
reflection processes will strongly depend on the position of the middle dot energy level $\varepsilon_{S}$. This is indeed the case: the Andreev conductance becomes
maximized when $\varepsilon_{S}=0$ and gets decreased with increasing the detuning of the middle dot level from the Fermi energy \cite{Dominguez2016}. However, it
turns out that while the total conductance strongly depends on $\varepsilon_{S}$, the ratio of the contributions due to DAR and CAR processes does not (not shown).
This is almost strictly obeyed in the linear response regime, while in the nonlinear response regime it holds for relatively low voltages. The same happens for the
dependence of the total Andreev current on magnetic configuration of the device, while the currents in both parallel and antiparallel configurations strongly depend on
the position of $\varepsilon_{S}$, their ratio only very weakly does so. Consequently, the TMR and the Cooper pair splitting efficiency, the quantities that are of
main interest in this paper, do not exhibit considerable dependence on $\varepsilon_{S}$. Therefore, we assume that the middle dot level is fixed during our analysis
and equal to $\varepsilon_{S}=-\Gamma/2$. In fact, this choice is also motivated by the work of F\"up\"ol {\it et al.} \cite{Fulop_PhysRevLett.115.227003}. In this
paper, the three-site model was introduced to describe the transport properties of quantum-dot-based Cooper pair splitters and it was shown that the best agreement
between theoretical modeling and experimental data was obtained for slightly detuned level position of the middle dot. }

In the following, we analyze the Andreev reflection transport properties of triple quantum-dot-based Cooper pair splitters for various parameters of the model. In
particular, we thoroughly discuss the behavior of the Andreev conductance, splitting efficiency and the tunnel magnetoresistance, both in the linear and nonlinear
response regimes. Let us start the discussion with the case of the linear response regime.

%~~~~~~~~~~~~~~~~~~~~~~~~~~~~~~~~~~~~~~~~~~~~~~~~~~~~~~~~~~~~~~~~~~~~~~~~~~~~~~~~~~~~~~~~~~~~~~~~~~~%
\section{Linear response regime}
\label{Sec:III}
%~~~~~~~~~~~~~~~~~~~~~~~~~~~~~~~~~~~~~~~~~~~~~~~~~~~~~~~~~~~~~~~~~~~~~~~~~~~~~~~~~~~~~~~~~~~~~~~~~~~%

\subsection{Andreev linear conductance}
\label{Sec:IIIA}

\begin{figure}[t]
    \includegraphics[width=0.95\columnwidth]{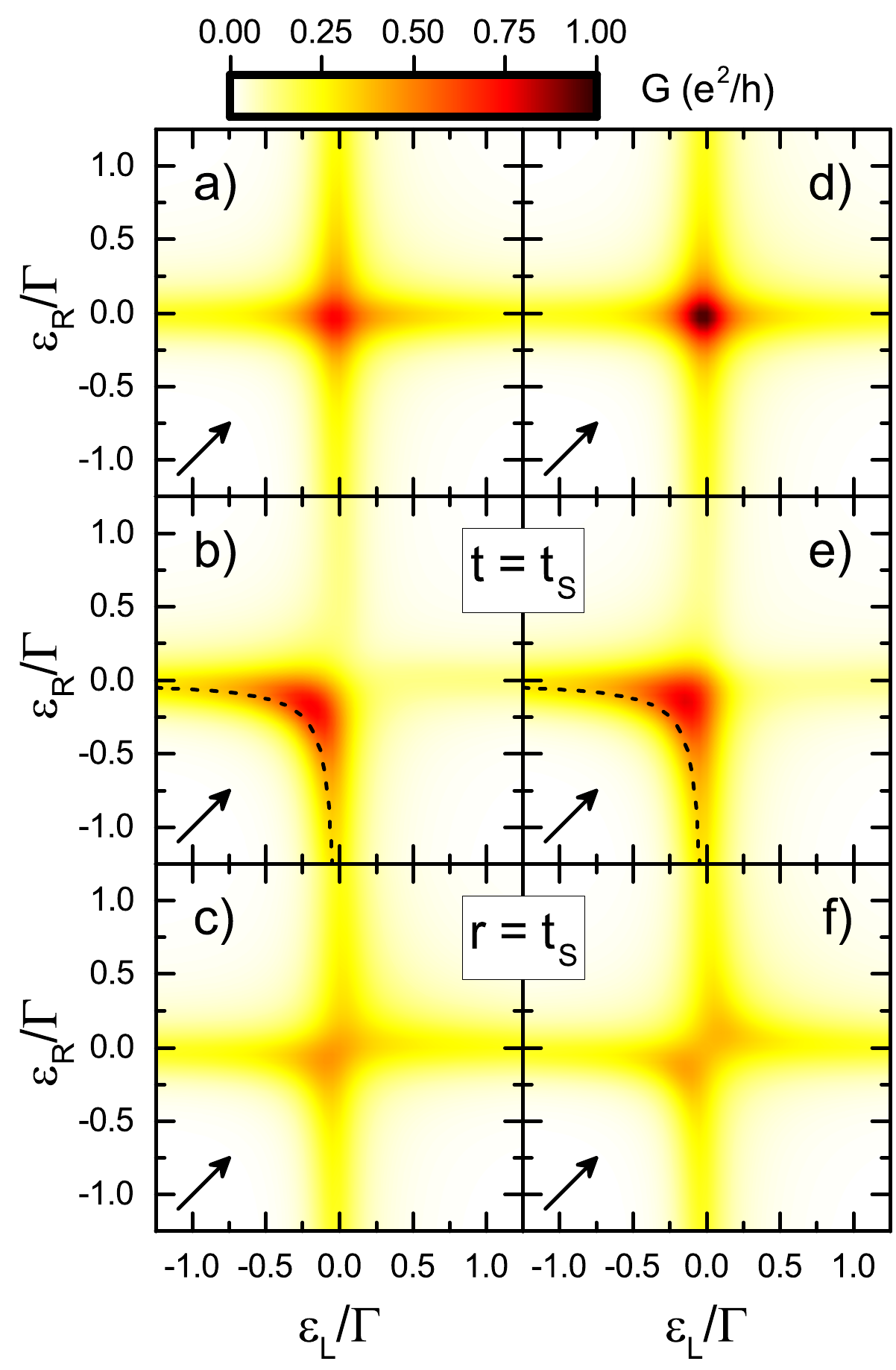}
    \caption{
        The total linear conductance due to Andreev reflection processes
        calculated as a function of the position of the left ($\varepsilon_L$)
        and right ($\varepsilon_R$) quantum dot levels.
        The left (right) column corresponds to the case of the parallel (antiparallel)
        magnetic configuration.
        The first row shows the results for $t=r=0$,
        the second one for $t=t_S$, $r=0$,
        whereas the third one is calculated for $t=0$, $r=t_S$.
        The other parameters are: $U=10\Gamma$, $\varepsilon_{S}=-\Gamma/2$,
        $\Gamma_{S}=2\Gamma$, $t_S=\Gamma/4$,
        with $\Gamma$ used as energy unit,
        and $p=0.5$.
        The dashed lines in (b) and (e) mark
        approximately the energy of the bonding state.
        Arrows indicate the cross-sections discussed in Fig.~\ref{Fig:G_along_diagonal}.
        \new{All calculations performed assuming
        the infinite superconducting gap limit.}
    }
    \label{Fig:G2D}
\end{figure}

The linear conductance calculated as a function of the
position of the left and right quantum dots' energy levels is shown in \fig{Fig:G2D}.
The left column displays the results for the parallel magnetic configuration,
while the right column presents the data for the antiparallel configuration.
The first row is calculated for the case when there is no direct coupling
between the left and right dots, while the second and third rows
correspond to the situation when the hopping is finite
and it is either normal or Rashba-type of hopping, respectively.

\begin{figure}[t]
    \includegraphics[width=1\columnwidth]{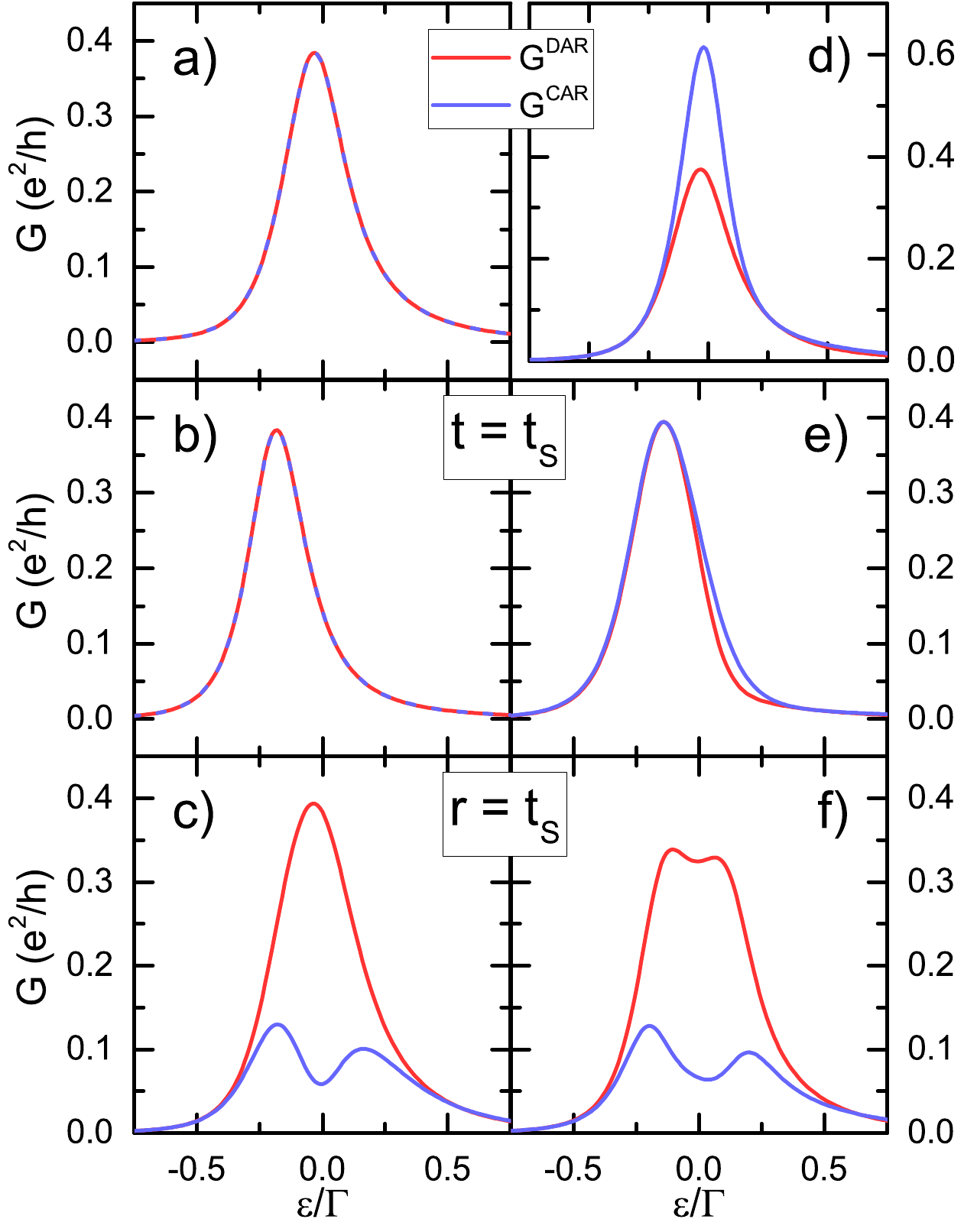}
    \caption{
        The contributions, $G^{\rm DAR}$ and $G^{\rm CAR}$,
        to the total linear Andreev conductance due to DAR and CAR processes,
        respectively, plotted as a function of the left and right dots' energy levels
        $\e_L=\e_R\equiv \e$.
        The first row is calculated for $t=r=0$,
        the second one for $t=t_S$, $r=0$,
        whereas the third row corresponds to the case of $t=0$, $r=t_S$.
        The left (right) column presents the conductance in the case of parallel (antiparallel)
        magnetic configuration of the device.
        The other parameters are the same as in \fig{Fig:G2D}.
        \new{Notice different scales of the vertical axes in (a) and (d).}
    }
    \label{Fig:G_along_diagonal}
\end{figure}

First of all, one can generally observe that the conductance in the antiparallel configuration is larger than that in the parallel configuration. This is associated
with the fact that in the latter configuration the minority spin channel results in the suppression of CAR processes, as compared to the antiparallel configuration
where a fast majority nonlocal channel dominates transport. Although there are quantitative differences in the behavior of the total conductance in the two magnetic
configurations, its qualitative behavior rather weakly depends on alignment of magnetic moments of the leads (see \fig{Fig:G2D}). Therefore, let us for the moment
focus on the general dependence of the total linear conductance, while the difference in magnetic configurations will be revealed when discussing the behavior of
specific components to the total conductance due to DAR and CAR processes, which are depicted in \fig{Fig:G_along_diagonal}.

As can be seen in \fig{Fig:G2D}, the Andreev conductance reaches the highest values when the energy levels of the left and right dots are around the Fermi energy,
$\e_L\approx \e_R\approx 0$. Then, both DAR and CAR processes are maximized. In the absence of direct hopping between the left and right dots, the conductance exhibits
a cross-like structure as a function of the positions of the corresponding dots' levels, [see \figs{Fig:G2D}(a) and (d)]. For $\e_L\approx \e_R\approx 0$, DAR
processes through the two arms of the splitter and nonlocal CAR processes are allowed, which results in an enhancement of the Andreev conductance. When detuning the
system from this special point, the rate of Andreev reflection processes becomes suppressed, such that both $G_{\rm P}$ and $G_{\rm AP}$ suddenly drop. This happens in
the whole $\e_L-\e_R$ plane considered in \fig{Fig:G2D}, i.e., the larger the detuning the smaller the conductance, except for the lines given by $\e_L\approx 0$ and
$\e_R\approx 0$. Along those lines, an enhanced conductance comes mostly from DAR contributions through the dot, the level of which is aligned with the Fermi energy.

The behavior of the conductance changes when there is a direct coupling between the left and right dots. In the case of normal hopping, the bonding and antibonding
states form between the two dots, which greatly affects the conductance, as can be seen in \figs{Fig:G2D}(b) and (e). For a two-level system, the energy of the bonding
state crosses the Fermi energy when $\e_L\e_R=t^2$. Although for our proximitized triple quantum dot system this is a crude approximation, one can see that the
behavior of the conductance can be already quite reasonably explained by invoking the above energy dependence of the bonding state, which is marked in
\figs{Fig:G2D}(b) and (e) with a dashed line. Clearly, there is a strong asymmetry in $G_{\rm P/AP}$ with respect to $\e_L+\e_R=0$: the largest conductance can be seen
along the line given approximately by the energy of the bonding state $\e_L\e_R=t^2$ for $\e_L,\e_R<0$. We note that the impact of the formation of bonding and
antibonding states on the rate of Andreev reflection processes is larger for CAR than for DAR processes, since direct Andreev reflection occurs through only one arm of
the device. This can be seen in \figs{Fig:G2D}(b) and (e), where the cross-like structure of the conductance, and especially the long tails for $\e_L\approx 0$ and
$\e_R\approx 0$, which are mainly due to DAR processes, are still clearly visible.

When the coupling between the left and right dots is of Rashba type, [see \figs{Fig:G2D}(c) and (f)], the maximum value of the total Andreev conductance decreases and
it is approximately two times smaller compared to the other cases discussed. However, the general qualitative behavior of the linear Andreev conductance is quite
similar to the case in the absence of direct hopping between the left and right dots. Smaller values of the total conductance are caused by the suppression of CAR
processes by the Rashba spin-orbit interaction, as can be explicitly seen in \figs{Fig:G_along_diagonal}(c) and (f).

We also note that a small asymmetry of the conductance plots
with respect to $\e_L+\e_R=0$ visible in \fig{Fig:G2D}(a)
is mainly caused by the displacement of the middle
dot energy level $\varepsilon_{S}$ from the Fermi energy.
This asymmetry is further enhanced in the case of finite $t$
[\figs{Fig:G2D}(b) and (e)] due to the formation of bonding and antibonding states.

% discussion of cross-sections
As already mentioned, the difference between the two magnetic configurations
becomes revealed when studying the behavior of separate contributions to the total conductance,
$G = G^{\rm DAR}+G^{\rm CAR}$.
These contributions are shown in \fig{Fig:G_along_diagonal}, where they are plotted as a function of the position of the left and right dots' levels,
$\varepsilon_{L}=\varepsilon_{R}\equiv\varepsilon$. The curves correspond to the cross-section of the conductance presented in \fig{Fig:G2D}, which is marked with an
arrow. Consider first the parallel magnetic configuration. In the absence of hopping between the left and right quantum dots, all contributions are equal [see
\fig{Fig:G_along_diagonal}(a)], which results in the splitting efficiency $\eta = 50\%$.
In fact, in this case one finds,
$G_{j\sigma}^{\rm DAR} = G_{j\sigma}^{\rm CAR} = G/8$.
This behavior can be understood
by realizing that the generation of Cooper pairs
is mainly governed by the minority bands of the ferromagnetic leads.
If there is a depletion of minority electron states,
the available majority electrons cannot form new Cooper pairs,
which results in suppression of the Andreev transport.
Because in the case of parallel configuration
the majority and minority bands of both leads are equal,
contributions from the local and nonlocal Cooper
pair tunneling processes are also the same.

If there is a finite hopping between the dots [\fig{Fig:G_along_diagonal}(b)], the conductance behavior does not change qualitatively too much. All contributions to
the total conductance are again equal due to the symmetry of the leads and their bands. The only difference is associated with the shift of the maximum in $G$ toward
lower energies of about the value of the interdot hopping $t$, which is due to the formation of bonding and antibonding states. On the other hand, in the case of
finite Rashba spin-orbit coupling between the left and right dots [\fig{Fig:G_along_diagonal}(c)], the CAR and DAR processes do not contribute equally any more. In
this situation, we find the following relations for the conductance components $G_{L\sigma}^{\rm DAR/CAR}=G_{R\sigma}^{\rm DAR/CAR}$. However, as can be seen in
\fig{Fig:G_along_diagonal}(c), there is a large difference between conductances due to DAR and CAR processes. It turns out that while DAR processes are weakly
dependent on the Rashba spin-orbit coupling [cf. \figs{Fig:G_along_diagonal}(a) and (c)] CAR processes become suppressed by the Rashba interaction, such that $G^{\rm
DAR} > G^{\rm CAR}$.
\new{
The suppression of CAR processes is especially effective for $\e\approx 0$. It is seen that detuning of the dots' discrete levels from the Fermi energy of the leads
gives rise to a double-peak structure of the CAR resonance with a local maxima at $\varepsilon \approx \pm r$, separated by a local minimum at  $\e= 0$ [see
\fig{Fig:G_along_diagonal}(c)].} Note also that for larger values of $\e$, i.e., $|\e| > r$, the DAR and CAR conductances become comparable.

The right column of \fig{Fig:G_along_diagonal}
corresponds to the antiparallel magnetic configuration of the device.
In this case, the conductance contributions
fulfill the following relation, $G_{L\sigma}^{\rm DAR/CAR}=G_{R\bar{\sigma}}^{\rm DAR/CAR}$,
which is simply related to the fact that the
spin subbands of the right lead are now flipped.
In addition, in the absence of Rashba spin-orbit coupling,
DAR contributions, which hardly depend on magnetic
configuration of the device, are all equal,
\new{
i.e., $G_{j\sigma}^{\rm DAR} = G^{\rm DAR}/4$. }
The conductance in the case of $t=r=0$ plotted as a function of $\e$ is shown in \fig{Fig:G_along_diagonal}(d). Clearly, DAR contributions remain independent of the
nonlocal change of polarizations of the leads resulting in almost unaltered behavior compared to that shown in \fig{Fig:G_along_diagonal}(a). Because in the
antiparallel configuration the majority bands of the leads have opposite spin directions, forming of the Cooper pairs is more efficient. As a result, there is a fast
CAR majority-spin channel, which gives the dominant contribution to the total conductance. Consequently, the crossed Andreev conductance around the Fermi level is much
enhanced with respect to the DAR conductance, which gives rise to high splitting efficiency, as will be discussed later on.

Interestingly, in the presence of direct hopping between the left and right dots, the ratio of DAR and CAR processes becomes modified. Now, this ratio strongly depends
on the amplitude of hopping $t$, contrary to the case of parallel configuration, where both processes contribute on an equal footing [cf. Figs.
\ref{Fig:G_along_diagonal}(b) and (e)]. Increasing the value of hopping results in the suppression of CAR processes, which leads to a reduced total conductance. We
note that for the selected value of hopping, the contributions from CAR and DAR processes are in fact comparable. However, further increase of $t$ causes the CAR
processes to dominate transport again. This implies that a direct hopping between the dots forming the arms of the splitter has a relatively strong effect on the
splitting efficiency depending on the value of the hopping, as will be shown in further sections.

When the coupling between the left and right dots is of Rashba type, the contribution to the conductance from CAR processes is much smaller compared to that due to DAR
processes. This situation is similar to the case of parallel configuration [see Figs. \ref{Fig:G_along_diagonal}(c) and (f)]. We note that in addition to a large
difference between DAR and CAR processes, there is also an imbalance between corresponding spin contributions flowing through given junctions (not shown). This is
quite counterintuitive as far as DAR processes are concerned, since one could expect that direct Andreev reflection, which occurs through one arm of the splitter,
should not depend on mutual orientation of magnetic moments of the left and right ferromagnetic leads. This is exactly the conclusion from the inspection of the
Andreev transport behavior in the absence of hopping between the dots and in the case of a finite direct hopping $t$. In the case of nonzero Rashba spin-orbit coupling
$r$, the situation is completely different. Spin-orbit interaction allows for a spin rotation during a hopping process, which in turn has an impact on the spin
dependence of processes through the left and right arms of the device. In terms of analytical formulas, finite spin-orbit coupling results in a new term in the Andreev
transmission coefficient, which is proportional to a product of two couplings corresponding to the same spin orientation [cf. Eqs. (\ref{eq:TDAR}) and
(\ref{eq:TCAR})]. Thus, for DAR processes, the spin channel proportional to majority-spin subband is favored, while in the case of CAR processes, the
spin-orbit-induced contribution is always proportional to a product of minority- and majority-spin-resolved couplings. This also explains the observed difference
between the magnitude of CAR and DAR processes seen in \fig{Fig:G_along_diagonal}(f).

\begin{figure}[t]
    \includegraphics[width=1\columnwidth]{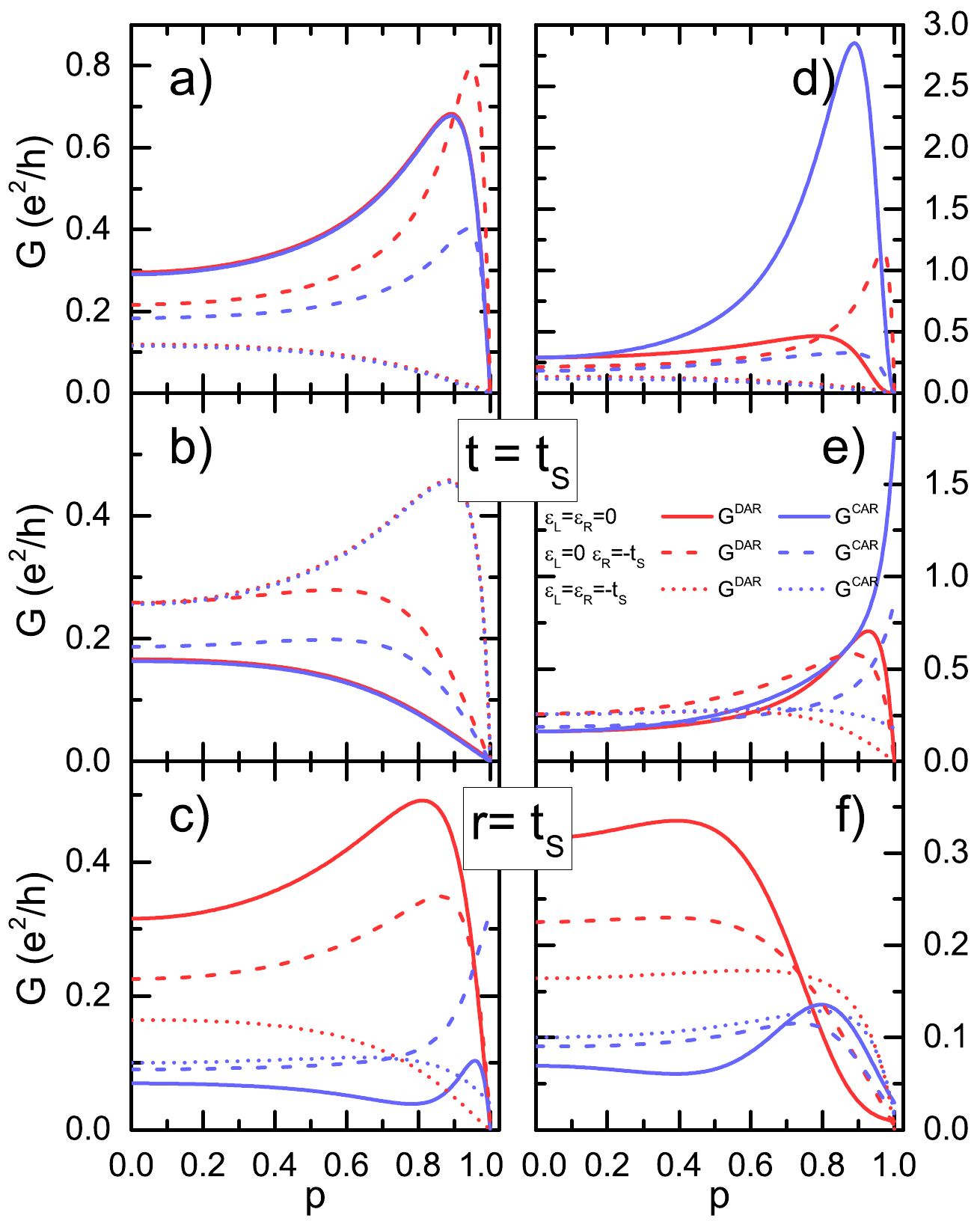}
    \caption{
        The Andreev conductance due to DAR and CAR processes
        plotted as a function of the spin polarization of the leads $p$
        in the parallel (left column) and antiparallel (right column) magnetic configuration.
        The first row is calculated for $t=r=0$,
        the second one for $t=t_S$, $r=0$,
        and the third row is for $t=0$, $r=t_S$.
        The conductance is plotted for selected
        values of $\e_L$ and $\e_R$, as indicated.
        The other parameters are the same as in \fig{Fig:G2D}.
        \new{Notice  different  scales of the vertical axes in the corresponding panels in the left and right column.}
    }
    \label{Fig:Gp}
\end{figure}

As follows from the above discussion, the rate of DAR and CAR processes
strongly depends on the position of the levels of the dots and the magnetic configuration of
the device. The difference between the local and nonlocal transport
events becomes even more enhanced when the spin polarization of the leads increases. This is
explicitly shown in \fig{Fig:Gp}, which presents the contributions
to the conductance due to DAR and CAR processes plotted as a function of spin polarization $p$ for
selected values of the dots' levels position. One can generally expect
an enhancement of CAR processes when the magnetic configuration changes from the parallel to the
antiparallel one, since then the nonlocal processes are mainly
determined by a fast majority spin channel, while the local processes are slower because they depend on
the minority spin subband \cite{Weymann2014Mar,Trocha2015Jun}.
However, it turns out that this rule can depend on the position of dots' levels and interference effects
\cite{Zhu_PRB01,Bai2012, Trocha2014_PhysRevB.89.245418,PSSB:PSSB201600206,Gong2016,Dominguez2016}.
Quantum interference can in fact make the situation quite
counterintuitive and reverse the role of CAR and DAR processes, as we discuss below.

Let us consider the case of parallel magnetic configuration.
For the first set of levels' position, $\e_L = \e_R = 0$, one can see that both DAR and CAR contributions
are equal, except for the case when there is Rashba interaction in action.
With increasing the spin polarization,
one observes a nonmonotonic dependence of $G^{\rm CAR/DAR}$ on $p$
for $t=r=0$ [\fig{Fig:Gp}(a)] with the following clearly visible features.
First, with raising $p$ to $p\approx 0.9$, the conductance increases, to become suddenly suppressed with further
increase of $p$. The vanishing of $G_{\rm P}$ in the limit of half-metallic leads can be easily understood.
When $p \to 1$, there is only one spin species and one of
the electrons forming the Cooper pair cannot tunnel
to the leads since there are no available states.
\new{
To explain such a spin polarization-dependent behavior of the conductance,
one should note that the magnitude of Andreev conductance
strongly depends on the ratio of the couplings to the normal electrodes and to the
superconductor \cite{Trocha2014_PhysRevB.89.245418,Zhu_PRB01,PhysRevB.88.155425}.}
When the spin polarization increases, this ratio is varied and so is the total Andreev conductance. For assumed parameters, increasing $p$ leads first to an
enhancement of the conductance. However, for large values of the spin polarization, the fact that the minority-spin subband is the bottleneck for Andreev processes
becomes relevant and eventually the dependence is changed, such that the conductance suddenly drops to get fully suppressed for $p=1$.

A similar dependence can be also observed in the case of $t=0$ and $r=t_S$
shown in \fig{Fig:Gp}(c), where additionally a large difference
between DAR and CAR processes occurs due to the Rashba spin-orbit coupling.
\new{
On the other hand, when the normal hopping between the dots is allowed, one observes a gradual decrease of $G$ with increasing $p$.
}%
This can be explained by realizing that now the conductance is generally low, since the maximum has been shifted to $\e_L = \e_R \approx -t$. In fact, one can observe
a similarity between the two cases of $t=r=0$ for $\e_L = \e_R = 0$ ($\e_L = \e_R = -t$ ) and $t=t_S$ ($r=0$) for $\e_L = \e_R = -t$ ($\e_L = \e_R = 0$). In the former
(latter) case, a nonmonotonic (monotonic) dependence of conductance on spin polarization is found [see Figs. \ref{Fig:Gp}(a) and (b)].
When the system is detuned in the $(\e_L,\e_R)$ parameter space
along the line $\e_L=0$ to $\e_R = -t$,
a qualitatively similar behavior to that discussed above can be seen.
However, now due to appropriate detuning,
the role of CAR processes is generally diminished,
and the Andreev conductance is mainly due to DAR processes.

\new{
A similar spin polarization-dependent behavior of the conductance
to that displayed in \fig{Fig:Gp}(a) can be also observed
in the case of antiparallel magnetic configuration,
which is depicted in the right column of \fig{Fig:Gp}.
}%
In this configuration one can expect CAR processes to dominate transport and this is exactly the case for $t=r=0$ and $\e_L = \e_R = 0$ [see \fig{Fig:Gp}(d)]. Note
that in this situation $G_{\rm AP}$ exhibits a local maximum around $p\approx 0.9$ and then, for $p\to 1$, drops to a very low but finite value. The conductance
suppression is associated with destructive quantum interference between local and nonlocal Andreev processes \cite{Trocha2014_PhysRevB.89.245418}. This interference
can become greatly affected by the presence of finite hopping between the two dots.
\new{
In particular,
in the case of $t=t_S$ and $r=0$ shown in \fig{Fig:Gp}(e),
the CAR conductance increases with $p$
to a local maximum for $p= 1$,
where Andreev transport is exclusively due to
crossed Andreev reflection.
}%
However, the dominant role of CAR processes
in the antiparallel configuration is not always observed.
As can be seen in \fig{Fig:Gp}, the ratio of DAR and CAR processes
strongly depends on the positions of dots' levels,
the type of hopping between the dots (or its absence)
and the value of spin polarization.
Consequently, this ratio can be tuned by gate voltages applied to the dots.
Nevertheless, we would like to remind that, irrespective of what the ratio
between the local and nonlocal Andreev processes
for intermediate values of $p$ is,
in the limit of $p\to 1$ DAR processes are not allowed
and the conductance is exclusively due to crossed Andreev reflection events.

%%%%%%%%%%%%%%%%%%%%%%%%%%%%%%%
\subsection{Tunnel magnetoresistance}
\label{Sec:IIIB}
%%%%%%%%%%%%%%%%%%%%%%%%%%%%%%%

\begin{figure}[t]
    \includegraphics[width=0.8\columnwidth]{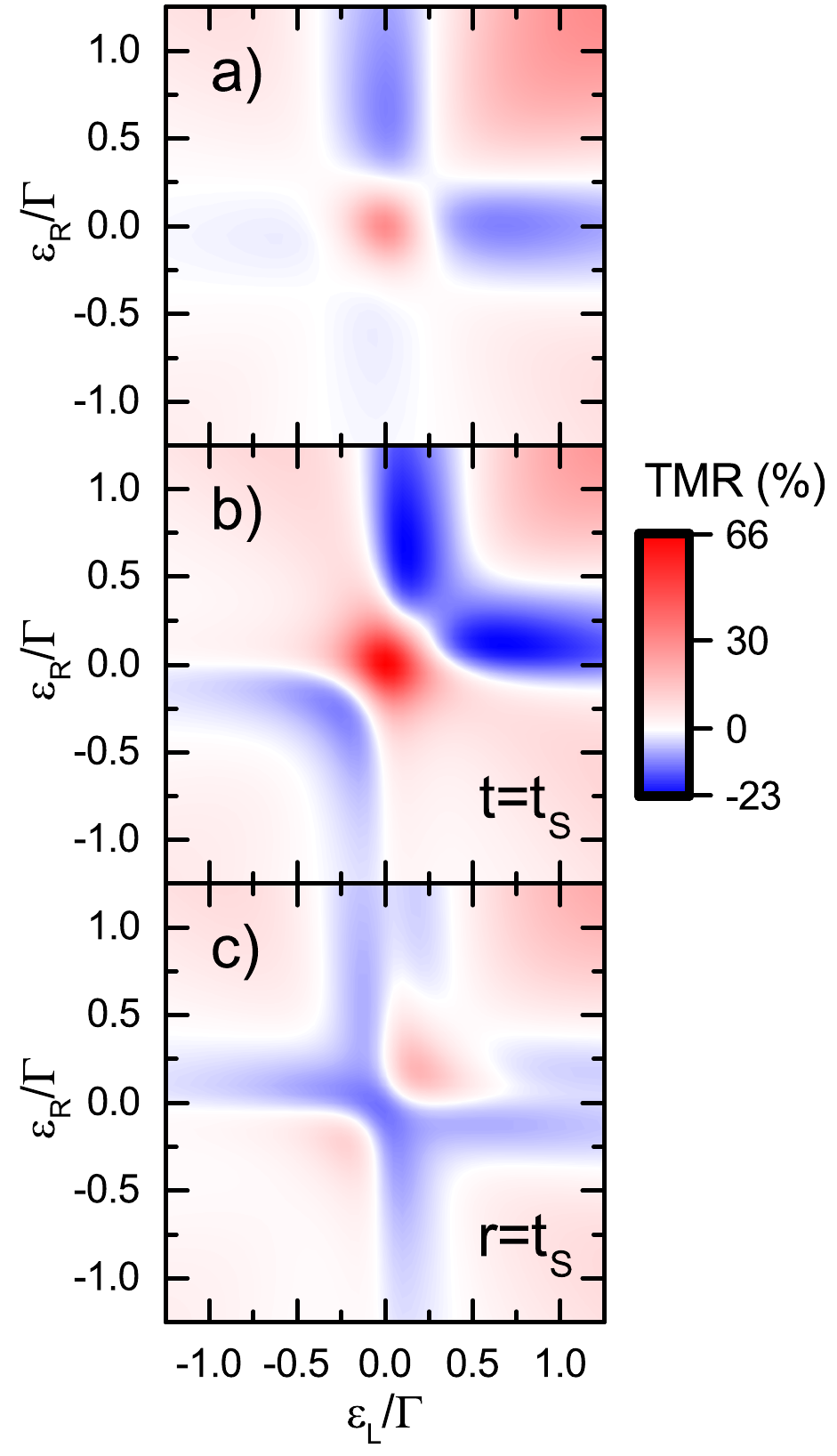}
    \caption{
        The tunnel magnetoresistance as a function of the level position
        of the left ($\varepsilon_{L}$) and right ($\varepsilon_{R}$) quantum dot
        calculated for: (a) $t=r=0$, (b) $t=t_S$, $r=0$ and (c) $t=0$, $r=t_S$.
        The other parameters are the same as in \fig{Fig:G2D}.
    }
    \label{Fig:TMR_equilibrium}
\end{figure}

To illustrate the quantitative differences between the Andreev transport behavior in the two magnetic configurations of the device in \fig{Fig:TMR_equilibrium} we show
the tunnel magnetoresistance calculated as a function of the quantum dots' levels $\varepsilon_{L}$ and $\varepsilon_{R}$. The TMR was in fact obtained from the
conductances shown in \fig{Fig:G2D} according to \eq{eq:TMR}. The colorscale is chosen in such a way that the blue (red) color corresponds to negative (positive) TMR.
As already mentioned, from the magnitude and sign of the TMR one can indirectly obtain some information about the mutual role of DAR and CAR processes in transport
\cite{Trocha2015Jun}. Large absolute values of TMR indicate that CAR processes are relevant, while suppressed TMR allows one to conclude that DAR processes are more important.
 Note, however, that in the presence of Rashba spin-orbit coupling one needs to analyze the data with an even greater care, since then DAR processes can
strongly depend on the system's magnetic configuration and it is not possible to draw such simple conclusions from the behavior of TMR about the role of DAR and CAR
processes.

Figure \ref{Fig:TMR_equilibrium}(a) presents the TMR calculated
in the situation without interdot hopping, $t=r=0$.
One can see that the largest TMR occurs around the Fermi level, $\varepsilon_{L}\approx \varepsilon_{R} \approx 0$,
which confirms that in the antiparallel configuration CAR processes
dominate transport. Along the lines when $\e_L\approx 0$
or $\e_R\approx 0$ and out of the Fermi level,
the TMR becomes negative, which in turn affirms a strong dependence of CAR processes on magnetic configuration of the device.
Interestingly, one can note that the asymmetry
along the line $\e_L + \e_R=0$ is now more
visible than in the conductance plots shown
in \figs{Fig:G2D}(a) and (b).
However, contrary to the conductance,
this asymmetry now hardly depends
on the position of the middle dot level,
and it mainly results from finite Coulomb correlations
and the fact that the levels of side quantum dots are around
the Fermi energy, i.e. they are detuned
from the particle-hole symmetry point of each dot.

The largest changes in the Andreev conductance when the magnetic configuration of the device is varied are found in the case of finite direct hopping amplitude $t$.
The TMR calculated for $t=t_{S}$ is presented in \fig{Fig:TMR_equilibrium}(b). It can be seen that for assumed parameters the TMR takes values ranging from ${\rm TMR}
\approx -23\%$ up to ${\rm TMR} \approx 66\%$. The largest values of the TMR are again observed around the Fermi energy, while negative TMR occurs for such energies
that $\e_L \e_R \approx t^2$, i.e., for the energies corresponding to the bonding (for $\e_L,\e_R<0$) and antibonding (for $\e_L,\e_R>0$) states [cf. \figs{Fig:G2D}(b)
and (e)]. Note that now the most negative TMR occurs for the antibonding state where, however, the conductance is smaller as compared to energies corresponding to the
bonding state.

On the other hand, in the case of finite Rashba spin-orbit coupling,
the values of the TMR are suppressed as compared
to the case with finite normal hopping [\fig{Fig:TMR_equilibrium}(c)].
One can see that now the TMR
takes values ranging approximately from
${\rm TMR}\approx -13\%$ to
${\rm TMR}\approx 20\%$.
The suppression of the TMR as compared to the case shown
in \fig{Fig:TMR_equilibrium}(b)
is associated with fact that
the spin-orbit coupling between the left and right dots
introduces a spin relaxation mechanism between the different spin channels
through the two arms of the splitter,
and it is a known fact that finite spin relaxation reduces the TMR
\cite{Rudzinski2001Aug}.
In the considered case, the spin-orbit coupling results thus in a reduction of the change
in transport properties when the magnetic configuration of the system is varied.
We also note that the main maximum around
$\varepsilon_{L}\approx \varepsilon_{R} \approx 0$, observed for $t=r=0$,
is now shifted towards higher energies by about
the value of the hopping amplitude $r=t_S = \Gamma/4$.

%%%%%%%%%%%%%%%%%%%%%%%%%%%%%%%
\subsection{Cooper pair splitting efficiency}
\label{Sec:IIIC}
%%%%%%%%%%%%%%%%%%%%%%%%%%%%%%%

\begin{figure}[t]
    \includegraphics[width=0.95\columnwidth]{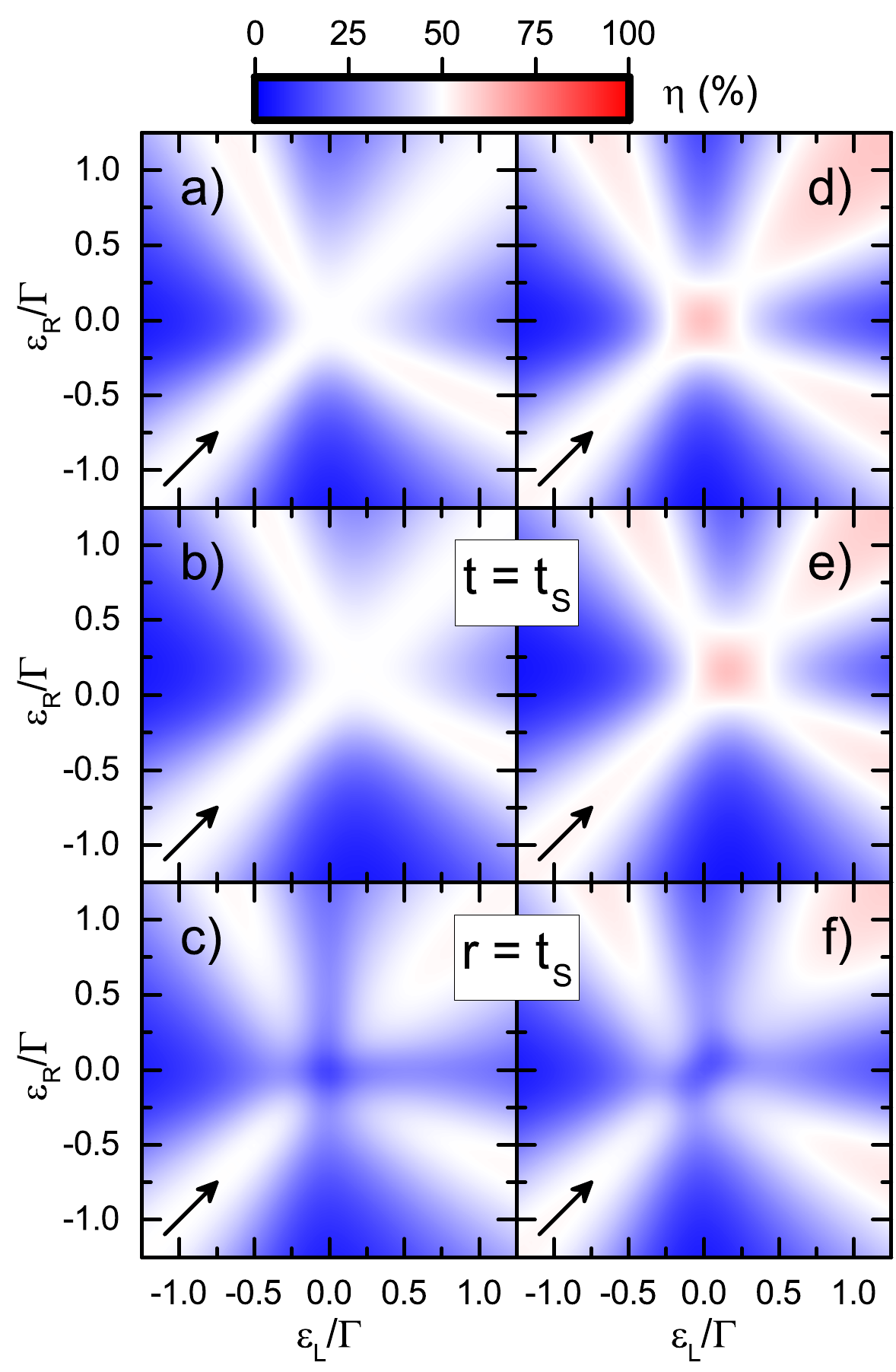}
    \caption{
        The Cooper pair splitting efficiency
        as a function of the position of the left ($\varepsilon_{L}$)
        and right ($\varepsilon_{R}$) quantum dots' energy levels
        calculated in the case of the parallel (a)-(c) and antiparallel (d)-(f) magnetic
        configuration of ferromagnetic leads.
        The first row corresponds to $t=r=0$, second
        row to $t=t_S$ and $r=0$, while the third row is for $t=0$ and $r=t_S$.
        Arrows indicate the cross-sections studied in detail
        in Fig.~\ref{Fig:Efficiency_vs_Ed_for_P}.
        The parameters are the same as in \fig{Fig:G2D}.
    }
    \label{Fig:Efficiency_equilibrium}
\end{figure}

Let us now discuss the behavior of one of the most important
quantities describing the ability of the device to split the Cooper pairs.
The splitting efficiency $\eta$ calculated as a function of the
position of the left and right dot energy levels is shown
in \fig{Fig:Efficiency_equilibrium} for the corresponding cases
as far as the coupling between the side dots is concerned.
The left (right) column corresponds to the case of parallel (antiparallel)
magnetic configuration.

First of all, one can see that in all the cases depicted in \fig{Fig:Efficiency_equilibrium} there is a pattern of enhanced splitting efficiency, which resembles a
cross-like structure, however, it is now rotated by $\pi/4$ with respect to the pattern visible in the behavior of the linear conductance shown in \fig{Fig:G2D}. From
this follows that a considerable splitting efficiency does not necessarily occur in transport regimes where the conductance is large. The enhanced efficiency occurs
along the following two branches: $\e_L \approx \e_R$ and $\e_L \approx -\e_R$ (see \fig{Fig:Efficiency_equilibrium}). The reason for this behavior can be understood
as follows. The CAR processes, in which the Cooper pair electrons are split into side dots are most effective when the energy levels are either aligned $\e_L \approx
\e_R$ or their average energy is equal to zero, which happens for $\e_L \approx -\e_R$. We also note that the branch along $\varepsilon_{L}\approx -\varepsilon_{R}$ is
slightly shifted toward the higher energies, which is associated with the Coulomb correlations. The second general observation is that the splitting efficiency is
lower in the case of parallel magnetic configuration compared to the antiparallel one. The enhancement of $\eta$ in the antiparallel configuration can be easily
explained as a result of an increased rate for crossed Andreev processes involving the spin-majority bands of both ferromagnetic electrodes.

\begin{figure}[t]
    \includegraphics[width=0.9\columnwidth]{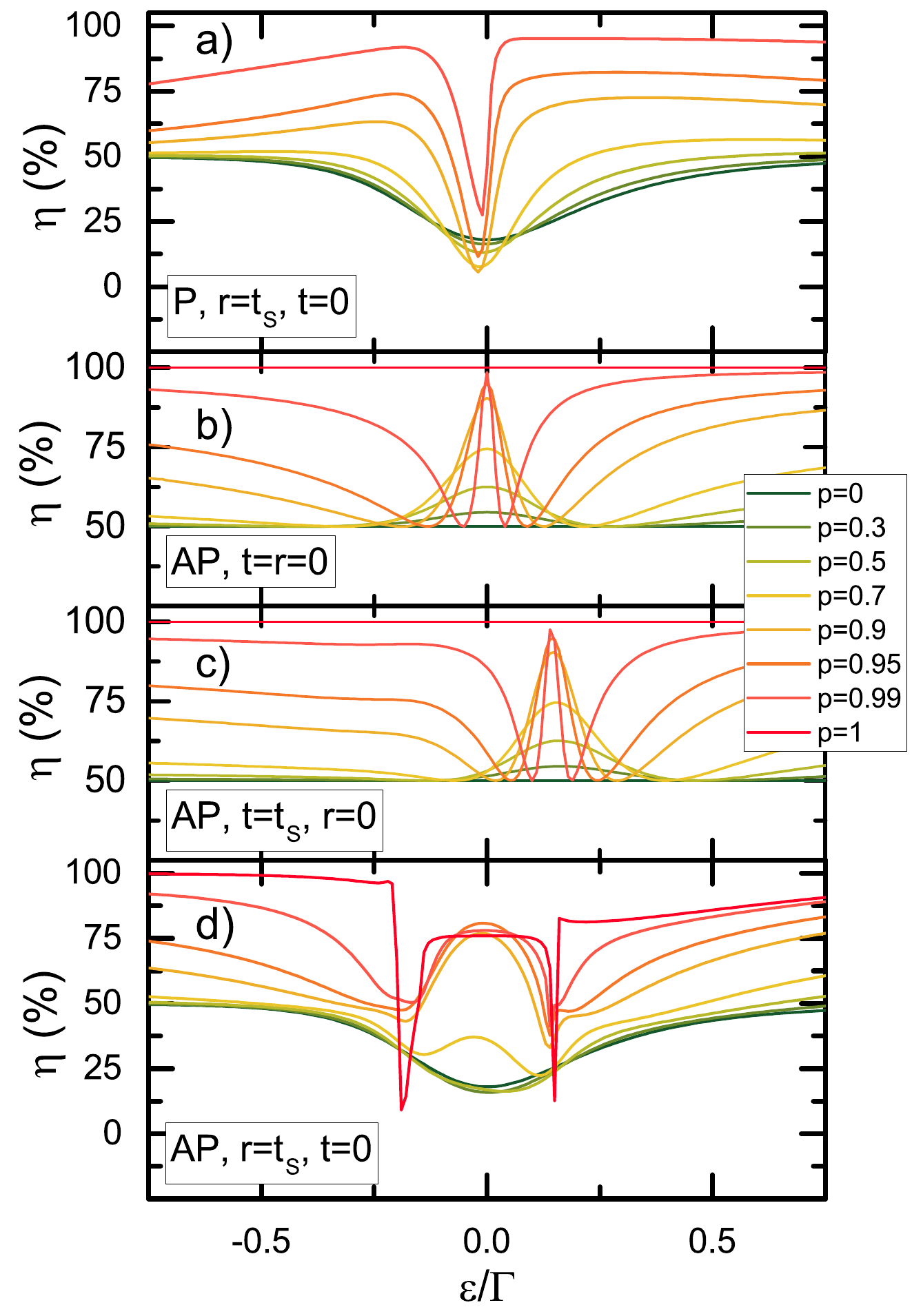}
    \caption{
        The efficiency $\eta$ of the Cooper pair splitting
        as a function of left and right dot energy levels $\varepsilon_{L}=\varepsilon_{R}=\varepsilon$
        calculated for selected values of leads' spin polarization $p$
        in the case of (a) parallel and (b)-(d) antiparallel magnetic configuration.
        The values of the hopping between the left and right dots
        are indicated in the figure, while other parameters are the same as in \fig{Fig:G2D}.
    }
    \label{Fig:Efficiency_vs_Ed_for_P}
\end{figure}

Let us now discuss the behavior of $\eta$ in the considered specific cases. It can be seen that, contrary to the conductance and TMR, the behavior of the splitting
efficiency hardly depends on the direct hopping $t$. In both cases, an enhanced efficiency occurs approximately along the diagonals in the $\e_L$-$\e_R$ plane. The
only difference is the shift of the cross-like structure, and thus the crossing point of the two branches, to the energy of antibonding state, $\varepsilon_{L}\approx
\varepsilon_{R}\approx t$. However, in contrast to the previous cases, when Rashba spin-orbit interaction is in action, the region of the high efficiency near the
Fermi level vanishes and the branches do not cross [see \figs{Fig:Efficiency_equilibrium}(c) and (f)]. This clearly demonstrates a detrimental impact of the spin-orbit
coupling between the left and right dots on CAR processes and, consequently, the splitting performance of the device.

In \fig{Fig:Efficiency_vs_Ed_for_P} we present the dependence of the splitting efficiency along the line given by $\varepsilon_{L}=\varepsilon_{R}=\varepsilon$, which
is marked with an arrow in \fig{Fig:Efficiency_equilibrium}. Because the rates of both CAR and DAR processes depend on the spin-resolved couplings, it is interesting
to analyze how $\eta$ depends on the spin polarization of ferromagnetic leads. In the case of parallel magnetic configuration, in the absence of spin-orbit coupling,
the splitting efficiency along the line $\varepsilon_{L}=\varepsilon_{R}=\varepsilon$ is equal to $50\%$ (not shown), irrespective of the value of the spin
polarization $p$. This is because tuning the spin polarization causes the majority and minority band couplings to change in the same fashion for both leads, such that
the ratio of DAR and CAR processes stays the same. Note, however, that when $p\to1$, i.e., for half-metallic leads, the Andreev reflection processes will be completely
suppressed, since there will be only one spin species in the ferromagnetic leads. In this case, the splitting efficiency will be an ill-defined quantity.

On the other hand, for the antiparallel magnetic configuration and for $r=0$, the growth of the spin polarization causes generally an enhancement of the splitting
efficiency [see \figs{Fig:Efficiency_vs_Ed_for_P}(b) and (c)], which grows from $\eta = 50\%$ to its maximum value. As already discussed above, enhancement of $\eta$
is a direct consequence of the fact that in the antiparallel configuration the minority spin channel is the bottleneck for DAR processes, while for CAR processes there
is always a fast majority-majority-spin channel. Consequently, in the limit of half-metallic leads, the splitting efficiency reaches its maximum value, since then only
crossed Andreev reflection processes are possible. Interestingly, one can note that the enhancement of $\eta$ with increasing the spin polarization $p$ exhibits a
strong dependence on the position of the energy levels of the left and right dots. Fast increase of $\eta$ with $p$ can be seen for $\e\approx 0$ in the case of
$r=t=0$, see \fig{Fig:Efficiency_vs_Ed_for_P}(b), however, there are also such values of $\e$ for which the splitting efficiency grows more slowly. In fact, $\eta$
displays a nonmonotonic dependence on $\e$, with a local minimum close to $\e \approx 0$. Such behavior is associated with interference between Andreev processes
involving two arms of the splitter. By comparing \figs{Fig:Efficiency_vs_Ed_for_P}(b) and (c) one can see that the dependence of $\eta$ is qualitatively very similar
in the case of $t=0$ and finite $t$, with the local maximum shifted approximately to the energy of the antibonding state [cf. \fig{Fig:Efficiency_equilibrium}(e)].

The situation, however, changes completely when there is a finite Rashba spin-orbit coupling between the side dots. For this case, the splitting efficiency is shown in
\figs{Fig:Efficiency_vs_Ed_for_P}(a) and (d) for the parallel and antiparallel magnetic configurations, respectively. First of all, one can note that even for
nonmagnetic leads the splitting efficiency is not equal to $50\%$, but becomes suppressed around $\e\approx 0$ and slowly grows with detuning the system from the Fermi
energy. This clearly demonstrates that finite spin-orbit coupling has a detrimental effect on the efficiency of the Cooper pair splitter. The splitting efficiency can
be enhanced by increasing the spin polarization of the leads, which can be seen for $p\gtrsim 0.5$. In the parallel configuration, $\eta$ generally grows with $p$
except for $\e\approx 0$ where a local minimum forms in the case of large spin polarization [see \fig{Fig:Efficiency_vs_Ed_for_P}(a)]. Interestingly, even in the case
of $p=1$ the conductance is not fully suppressed. This is because of the spin-orbit coupling which allows for flipping the spin of one of the split Cooper pair
electrons, such that the two electrons can tunnel either to the same or to different leads. This mechanism is also responsible for lowering of $\eta$ in the case of
antiparallel configuration for large spin polarizations [see \fig{Fig:Efficiency_vs_Ed_for_P}(d)]. Now, contrary to the cases shown in
\figs{Fig:Efficiency_vs_Ed_for_P}(b) and (c), $\eta$ does not reach $100\%$ for $p=1$, which indicates that the spin-orbit coupling results in finite DAR processes
even for half-metallic leads.

\begin{figure}[t!]
    \includegraphics[width=0.72\columnwidth]{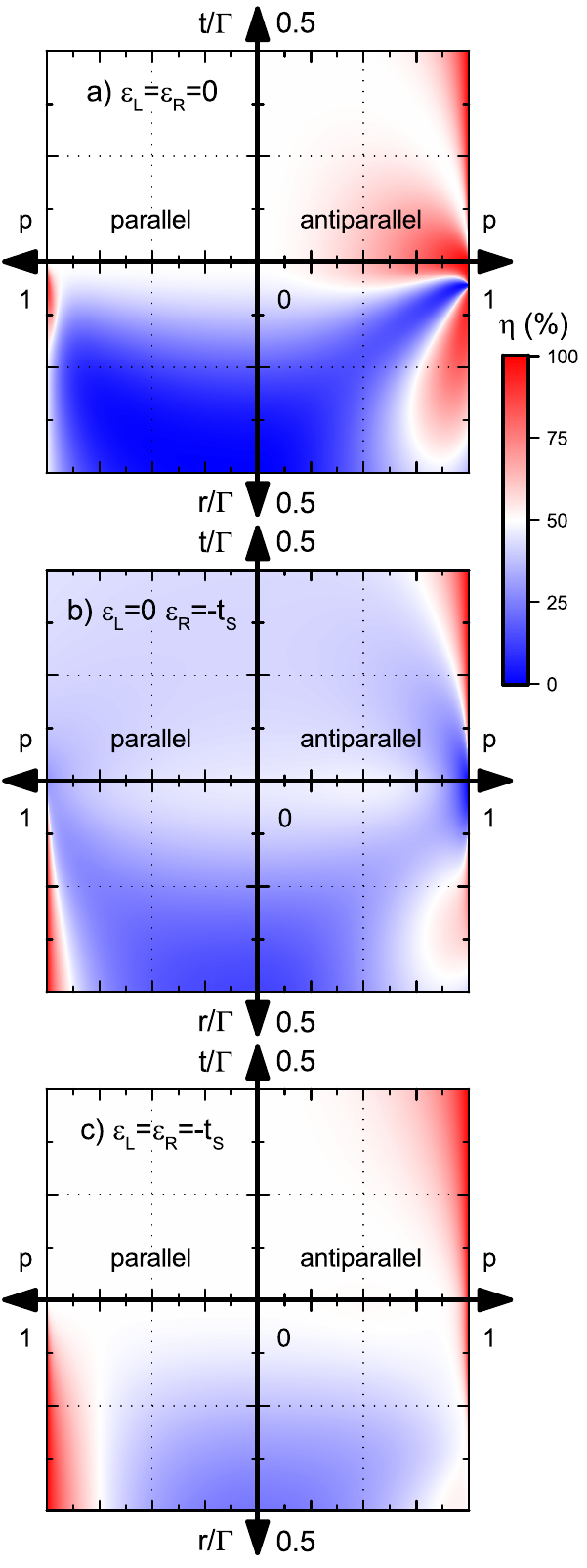}
    \caption{
        The efficiency of the Cooper pair splitting $\eta$
        as a function of the spin polarization of the leads $p$
        and the coupling strength between the dots, $t$ or $r$,
        calculated in the case of parallel and antiparallel configurations.
        Panel (a) shows $\eta$ calculated for $\varepsilon_{L}=\varepsilon_{R}=0$,
        (b) is determined for $\varepsilon_{L}=0$ and $\varepsilon_{R}=-t_S$,
        while (c) is for $\varepsilon_{L}=\varepsilon_{R}=-t_S$.
        The left (right) side of each panel corresponds
        to the parallel (antiparallel) configuration,
        while the upper (bottom) part shows the dependence
        on normal hopping $t$ (Rashba coupling $r$) and $p$.
        The crossing of the dotted lines specify the typical values used in this paper,
        i.e., $t=r=t_S$, with $t_S=\Gamma/4$, and $p=0.5$.
        The other parameters are the same as in \fig{Fig:G2D}.
    \vspace{-0.2cm}}
    \label{Fig:eta2D}
\end{figure}

Finally, we study the Cooper pair splitting efficiency by continuously changing the hopping between the dots and the spin polarization of the leads in both magnetic
configurations. The numerical results are displayed in \fig{Fig:eta2D}, where the intersections of the dotted lines correspond to $\eta$ calculated for typical values
of the parameters used throughout this paper, namely, $t=r=t_S$ ($t_S = \Gamma/4$) and $p=0.5$.

Let us first analyze the dependence of $\eta$ on the spin polarization of the leads $p$ and amplitude of direct hopping between the left and right dots $t$,
$\eta(p,t)$, for quantum dots' levels positions set at $\varepsilon_L=\varepsilon_R=0$ and $\varepsilon_L=\varepsilon_R=-t_S$, which is shown in \figs{Fig:eta2D}(a)
and (c), respectively. It can be clearly seen that, regardless of the values of the parameters $t$ and $p$, the splitting efficiency in the case of parallel
configuration equals $\eta \approx 50\%$, which follows from equal contributions of the CAR and DAR processes to the total conductance [see also the solid curves in
\figs{Fig:Gp}(a) and (b)]. This is, however, not the case in the antiparallel configuration when in the case of $\varepsilon_L=\varepsilon_R=0$ one observes a
significant enhancement of $\eta$ within two areas in the $(p,t)$ plane [see \fig{Fig:eta2D}(a)]. One such area extends approximately for $0\lesssim
{t}\lesssim{\Gamma/4}$ and $0\lesssim{p}\lesssim{1}$. In the presence of weak interdot coupling $t$, the splitting efficiency exhibits a maximum for large spin
polarization $p$, and then it drops for $p\to 1$. As discussed earlier [cf. the solid curves in \fig{Fig:Gp}(d)], this is due to the interference between the local and
nonlocal Andreev processes. The second area in the $(p,t)$ plane where $\eta$ is enhanced can be seen for relatively large spin polarizations and finite hopping $t$.
The region with $\eta>50\%$ emerges for $p\to1$ and $t\approx \Gamma/8$ and it spreads with increasing $t$ towards smaller values of spin polarizations. Thus, a
triangle-shaped area exhibiting large splitting efficiency is formed in the upper right corner of the $(p,t)$ plane shown in \fig{Fig:eta2D}(a). The observed behavior
of $\eta$ can be explained by realizing that with increasing the coupling amplitude $t$, the hopping processes amplify the CAR processes, thus reducing the negative
interference.
\new{
The latter occurs for large spin polarizations $0.8\lesssim p\lesssim 1$, when the CAR conductance starts to dominate over the DAR contribution [see the solid curves
in \fig{Fig:Gp}(e)].
}%
On the other hand, the dependence of $\eta$ on $p$ and $t$ shown in \fig{Fig:eta2D}(c)
reveals that even further extension of the discussed triangle-shaped region
with $\eta>50\%$ is possible if other tunings of the dots' discrete
levels are taken into account.
In particular, for $\varepsilon_L=\varepsilon_R=-t_S$,
a significant reduction of the Andreev processes
when the interdot coupling is weak occurs,
which is accompanied with $\eta\approx 50\%$.
Only for large enough spin polarization $p$
the hopping $t$ can significantly enhance the magnitude of CAR processes
[see also the dotted lines in \fig{Fig:Gp}(e)],
leading thus to a large splitting efficiency, $\eta\approx100\%$.

It is also worth to note that the influence of the interdot hopping
on the Cooper pair splitting efficiency in the antiparallel magnetic
configuration is still crucial when
the levels of the dots are detuned as, $\varepsilon_L=0, \varepsilon_R=-t_S$.
This is the case shown in \fig{Fig:eta2D}(b),
where it can be seen that even though an overall suppression
of the splitting efficiency occurs in both magnetic configurations,
there is a certain region in the $(p,t)$ plane, where $\eta>50\%$
for the antiparallel configuration case.

The impact of finite Rashba spin-orbit coupling $r$ between the left and right dots and the spin polarization $p$ on the splitting efficiency is presented in the
bottom parts of the panels shown in \fig{Fig:eta2D}. It is visible that in the case of parallel magnetic configuration the splitting efficiency becomes generally
suppressed $\eta < 50\%$ for all the level positions considered in the figure. Only when the spin polarization becomes relatively large ($0.8\lesssim p\lesssim 1$) one
can identify regions with $\eta>50\%$ in the ($p,r$) plane. More specifically, the largest such region is found for $\varepsilon_L=\varepsilon_R=-t_S$ and it is seen
in the left lower corner of the $(p,r)$ plane in \fig{Fig:eta2D}(c). The origin of this enhancement can be understood if one realizes that the spin-flip processes
during hopping between the dots effectively increase the number of spin states that can be occupied by the reflected carriers. As a consequence, this increases the
number of tunneling events that transfer entangled electrons into different external leads with parallel aligned magnetizations. By detuning the energy levels to
$\varepsilon_L=0$ and $\varepsilon_R=-t_S$ [\fig{Fig:eta2D}(b)], the DAR-type Andreev transmission dominates the current [see also the dashed lines in
\fig{Fig:Gp}(c)], which suppresses the splitting efficiency of the system. This tendency is even stronger if one tunes the dots' levels closer to the Fermi energy. In
particular, for the case of $\varepsilon_L=\varepsilon_R=0$ shown in \fig{Fig:eta2D}(a), one can see that the spin-orbit coupling completely suppresses CAR processes,
and also the region of enhanced splitting efficiency for $p\to 1$ vanishes almost entirely for the parallel magnetic configuration.

The spin-orbit coupling also strongly affects the CAR processes in the case of antiparallel alignment between the magnetizations of the ferromagnetic leads. Its
largest impact on the behavior of $\eta$ can be noticed for $\varepsilon_L=\varepsilon_R=0$. As can be seen in the lower right panel of \fig{Fig:eta2D}(a), one region
of enhanced splitting efficiency in the ($p,r$) plane is located for $0.25\lesssim p \lesssim {1}$ and for weak spin-orbit couplings,
$0\lesssim{r}\lesssim{\Gamma/16}$. Enhanced efficiency develops due to that fact that rare hoppings between the dots accompanied by spin rotation stimulate the CAR
contribution to the total conductance, especially when $p\to 1$. Moreover, an even larger area of enhanced splitting efficiency occurs for $0.5\lesssim{p}\lesssim{1}$
and $r\gtrsim \Gamma/16$. This enhancement can be understood if one compares the conductance characteristics with the corresponding splitting efficiency along the
horizontal dotted line in \fig{Fig:Gp}(f), from which it follows that for large enough polarizations $p\gtrsim 0.8$ the interplay between the CAR and DAR processes
enhances significantly the splitting efficiency. This effect is, however, diminished if a finite detuning of the level of one of the dots occurs [see
\fig{Fig:eta2D}(b)], and finally vanishes entirely in case when the levels of the both side quantum dots are shifted away from the Fermi level [see
\fig{Fig:eta2D}(c)]. However, it is also interesting to note that, simultaneously, for weak spin-orbit couplings, a small area with perfect splitting efficiency
$\eta\approx100\%$ may develop, provided that $p\approx1$.

%%%%%%%%%%%%%%%%%%%%%%%%%%%%%%%%%%%%%%%%%%%%%%
\section{Nonequilibrium regime}
\label{Sec:IV}
%%%%%%%%%%%%%%%%%%%%%%%%%%%%%%%%%%%%%%%%%%%%%%

In this section, we discuss the nonequilibrium Andreev transport properties of the considered device. We recall that the voltage is applied in such a way that the
superconductor is grounded, while the chemical potentials of both ferromagnetic leads are kept equal, $\mu_L = \mu_R = eV$.

%%%%%%%%%%%%%%%%%%%%%%%%%%%%%%%%%%%
\subsection{Differential Andreev conductance}
\label{Sec:IVA}
%%%%%%%%%%%%%%%%%%%%%%%%%%%%%%%%%%%

\begin{figure}[t]
    \includegraphics[width=1\columnwidth]{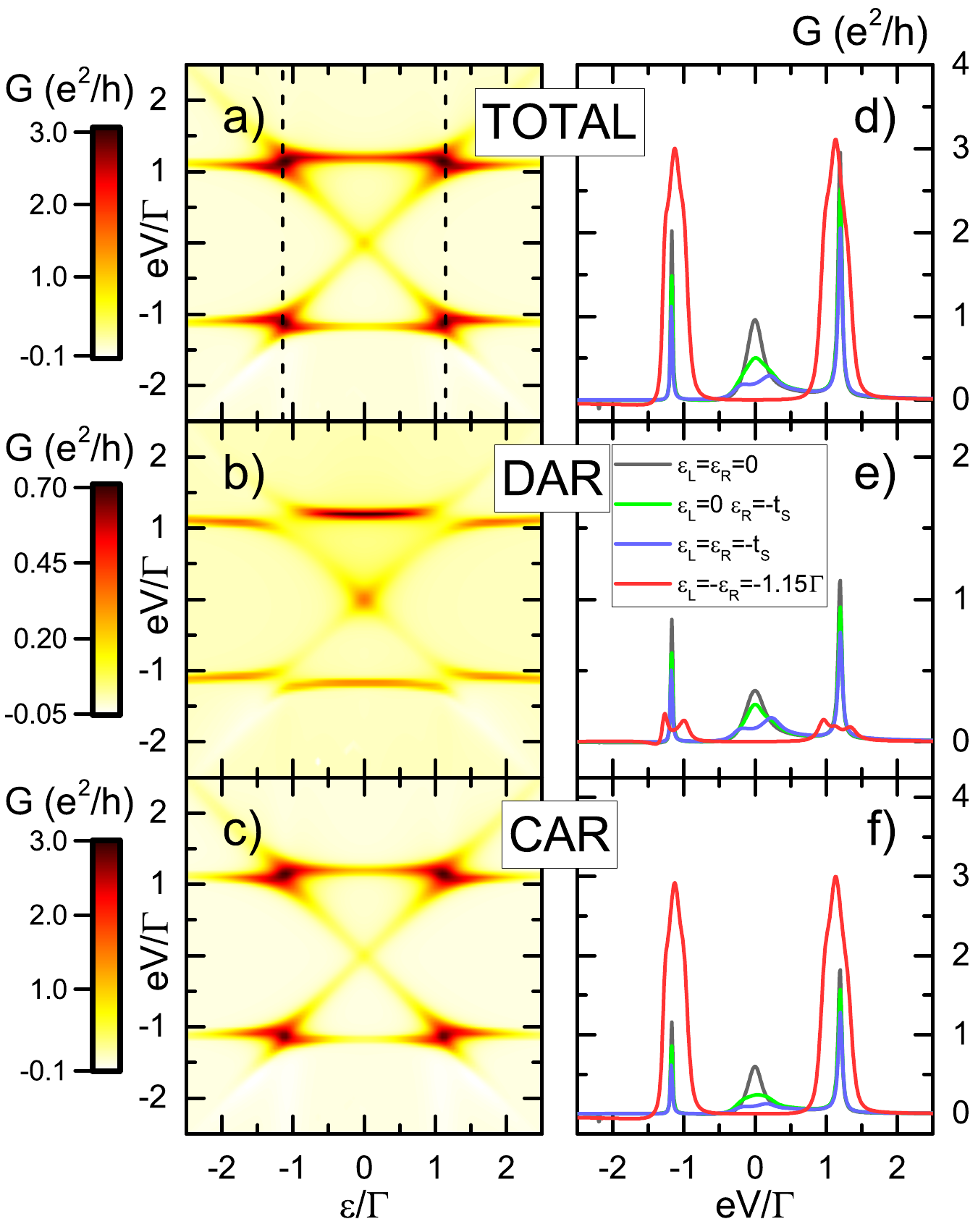}
    \caption{
       The differential Andreev conductance in the
      antiparallel magnetic configuration, without hopping between the dots ($t=r=0$), calculated as a function
      of the left and right quantum dots' levels $\e = \varepsilon_{L}=-\varepsilon_{R}$
     and applied bias voltage $V$.
     The left column shows the map of (a) the total  conductance, as well as (b) DAR and (c) CAR contributions.
     The right column shows the differential conductance
     as a function of the applied bias voltage for
     selected values of the dots' levels positions, as indicated in the legend.
     The other parameters are the same as in \fig{Fig:G2D}.
  }
 \label{Fig:Gdiff}
\end{figure}

In \fig{Fig:Gdiff} we present the differential Andreev conductance (the total one as well as CAR and DAR contributions) plotted as a function of the applied bias
voltage $V$ and the level position $\e$ with $\varepsilon=\varepsilon_L=-\varepsilon_R$. For clarity, we present here only the results calculated for the antiparallel
magnetic configuration and for a specific case of $t=r=0$, i.e., in the absence of hopping between the dots. In fact, the conductances obtained for the parallel and
antiparallel configurations exhibit rather subtle qualitative and quantitative differences, even if the interdot hopping or the Rashba coupling is switched on. The
phenomena arising due to the change of magnetic configuration of the device are better revealed when examining the tunnel magnetoresistance and the splitting
efficiency, which will be discussed in the next sections.

The results presented in \fig{Fig:Gdiff} show that, due to the proximity effect, the Andreev bound states enhance both the local and nonlocal transmissions through the
junction, giving rise to the conductance maxima. The main resonance peaks appear for $eV\approx \pm\varepsilon_A$, with
$\varepsilon_A=\sqrt{\varepsilon_{S}^{2}+\Gamma_{S}^{2}/4}$ describing approximately the Andreev bound-state energies \cite{Weymann2014Mar}. Moreover, one can note
that the height of the resonance peaks strongly depends on the relation between the energies of the dots' discrete levels relative to the energies of the Andreev bound
states. It is also found that to obtain the best transmission at the resonances, the condition $\varepsilon_L=-\varepsilon_R$ should be satisfied. This can be
explicitly seen in \figs{Fig:Gdiff}(d)--(f), where the respective cross sections of the maps presented in \figs{Fig:Gdiff}(a)--(c) are displayed. The bias-voltage
dependence in the case of $\varepsilon_L=\e_R=0$ and ${\varepsilon_L=\e_R=-1.15\Gamma \approx -\e_A}$, representing the cross sections along the dashed lines in
\fig{Fig:Gdiff}(a), clearly exhibits higher resonance peaks as compared to those found for other dot-level positions, such as $\varepsilon_L=0$, $\varepsilon_R=-t_S$
or $\varepsilon_L=\varepsilon_R=-t_S$, represented in \fig{Fig:Gdiff} by the green and blue lines, respectively. Furthermore, it can be also seen in \fig{Fig:Gdiff}
that in the vicinity of the threshold bias voltage, $eV\approx \pm{\e_A}$, the nonlocal Andreev processes are more effective compared to the local ones, and that the
largest difference between the CAR and DAR contributions occurs for $\varepsilon\approx \pm \e_A$.

Another interesting feature visible in \fig{Fig:Gdiff} is the asymmetry of the transport characteristics with respect to the bias reversal
\cite{Hussein2016,Trocha2015Jun}. This asymmetry follows from the interplay between the Coulomb correlations on the side quantum dots and the transport processes that
occur for positive and negative bias voltages. For the assumed value of onsite Coulomb repulsion on the side dots, $U=10\Gamma$, these dots may be either empty or
singly occupied in the range of bias voltages taken into account. When $eV>0$, then around the threshold voltage, $eV \approx \e_A$, the Andreev energy level becomes
activated and thus the extracted Cooper pairs give rise to an enhancement of the Andreev current and to the resonance maximum in the differential conductance. However,
upon the bias voltage reversal, $eV<0$, the situation changes such that now the Andreev level of energy $-\varepsilon_A$ enters the transport window and the carrier
transmission into the superconductor amplifies creation of Cooper pairs. Simultaneously, negative bias voltage enhances the occupancy of the side dots by electrons
tunneling from ferromagnetic leads. These processes suppress the Andreev backward transmission, thus giving rise to a lower resonance maximum of the differential
Andreev conductance at $eV \approx -\e_A$ [see \fig{Fig:Gdiff}(a)].

%%%%%%%%%%%%%%%%%%%%%%%%%%%%%%%%%%%
\subsection{Tunnel magnetoresistance}
\label{Sec:IVB}
%%%%%%%%%%%%%%%%%%%%%%%%%%%%%%%%%%%

\begin{figure}[t]
    \includegraphics[width=1\columnwidth]{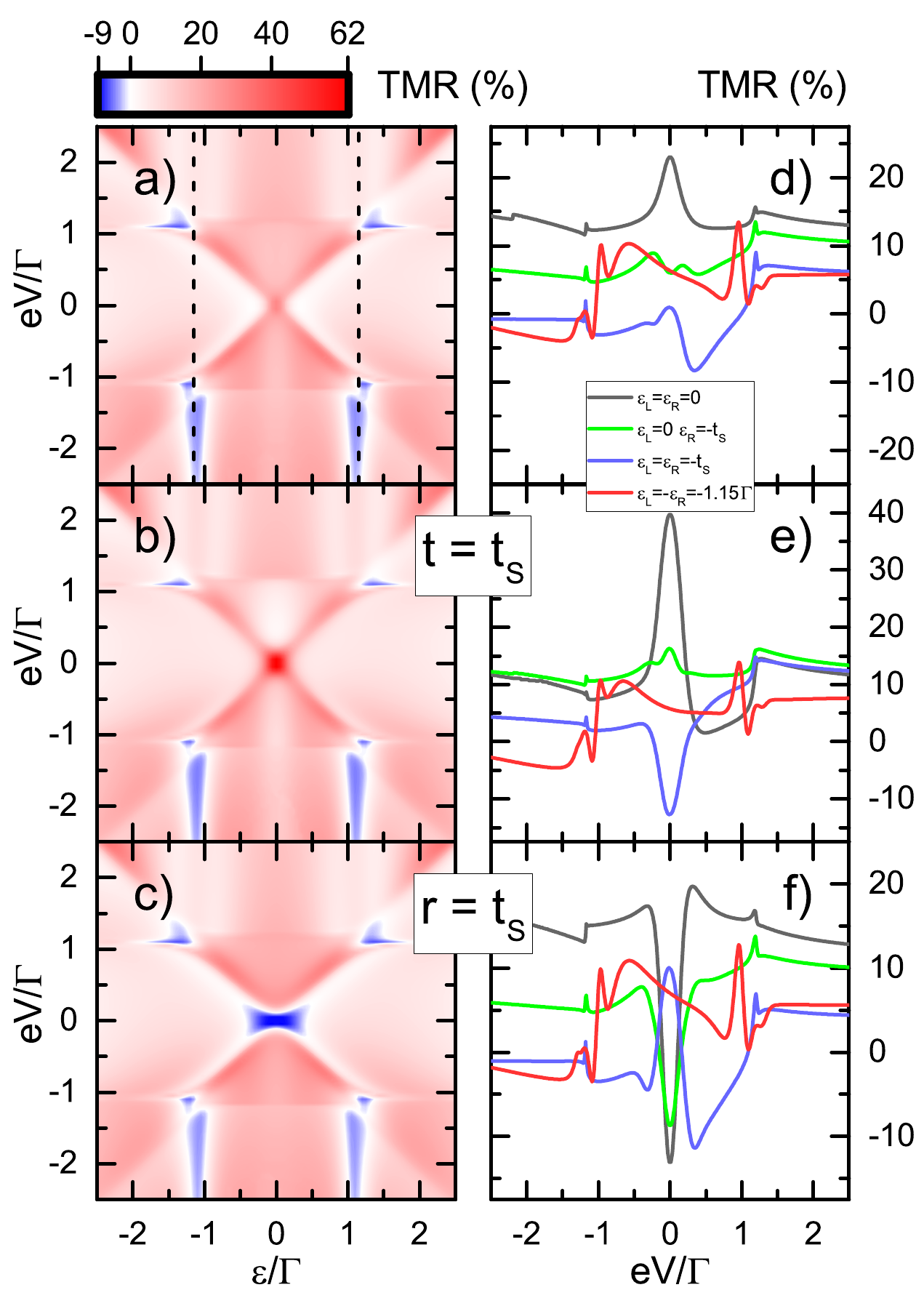}
    \caption{
        The tunnel magnetoresistance as a function of the position
        of the left and right quantum dot levels
        $\e = \varepsilon_{L}=-\varepsilon_{R}$ and
        the applied bias voltage $V$ (left column).
        The right column presents the
        bias dependence of the TMR
        for different dot level positions, as indicated.
        The first row corresponds to the case without hopping between the dots,
	while the second (third) row is calculated for
	$t=t_S$ and $r=0$ ($t=0$ and $r=t_S$).
	The other parameters are the same as in \fig{Fig:G2D}.
    }
    \label{Fig:TMR_Nonequilibrium}
\end{figure}

The dependence of the TMR on the bias voltage and the position of the dots' levels $\e = \varepsilon_{L}=-\varepsilon_{R}$ is shown in the left column of
\fig{Fig:TMR_Nonequilibrium}, while the right column displays the corresponding bias dependence of the TMR for selected values of $\varepsilon_{L}$ and
$\varepsilon_{R}$. As a first general observation, one can consider a strong dependence of the TMR on the magnitude of hopping between the dots for low voltages and
its absence when the applied voltage is larger than the hopping amplitude. The second observation is that the TMR is positive in almost the whole parameter space
considered in the figure, except for small regions where it can become negative (see \fig{Fig:TMR_Nonequilibrium}).

Let us now discuss in somewhat greater detail the nonequilibrium TMR behavior. In the case of $t=r=0$, which is displayed in \figs{Fig:TMR_Nonequilibrium}(a) and (d),
one can see that when the dots' levels are tuned to the energies corresponding to the Andreev bound states, ${\e_L=-\e_R = -1.15 \Gamma\approx -\e_A}$, the TMR
oscillates as a function of $eV$ with a rather small amplitude that ranges from ${\rm TMR}\approx -5\%$ to ${\rm TMR}\approx 15\%$ [see the red solid line in
\fig{Fig:TMR_Nonequilibrium}(d) corresponding to the vertical dashed lines in \fig{Fig:TMR_Nonequilibrium}(a)]. We have found that for negative bias voltage, the TMR
changes sign for $eV\lesssim -\e_A$, which is due to the fact that when the Andreev bound state energy level  $-\varepsilon_A$ enters the transport window, parallel
alignment of magnetizations of external leads gives rise to a significant amplification of the nonlocal transport processes. Moreover, as follows from
\figs{Fig:TMR_Nonequilibrium}(b), (c), (e), and (f), direct hopping or Rashba-type of coupling very weakly affects the transmission in both magnetic configurations as
long as the dots' energy levels are tuned to the Andreev bound state energies $\e_L = -\e_R \approx \pm{\varepsilon_A}$. This can be explained by realizing that there
are two different energy scales that can influence the Andreev reflection. The first one is associated with the formation of Andreev bound states and depends on
$\Gamma_S$, whereas the second one is the hopping amplitude, either $t$ or $r$. Since in our analysis we consider $\Gamma_S$ to be larger than the inter-dot hopping
amplitudes, the Andreev bound states are hardly affected by the magnitude of $t$ and $r$ quantities. As a consequence, when the energies of the dots are tuned to
$\varepsilon_L=-\varepsilon_R\approx \pm \varepsilon_A$, the nonequilibrium transport characteristics remain unmodified, regardless of the strength of direct coupling
between the dots (provided that $\Gamma_S$ is larger than the corresponding hoppings).

On the contrary, if the dots' levels are tuned to the Fermi energy, the Andreev transmission for low bias voltages exhibits a strong dependence on the strength and
type of coupling between the left and right dots. The bias dependence of the TMR for $\varepsilon=0$ is represented by the grey curves in
\figs{Fig:TMR_Nonequilibrium}(d)-(f). One can clearly see that for $t=r=0$ TMR displays a local maximum with ${\rm TMR}\approx25\%$ around the zero bias. This value
can be further enhanced up to ${\rm TMR}\approx40\%$ when the interdot coupling amplitude $t$ becomes finite [see \fig{Fig:TMR_Nonequilibrium}(e)]. Furthermore, the
TMR modification is even more prominent in the case of hopping accompanied by a spin flip. However, in such a case DAR processes dominate transmission, so that the
Andreev current in the parallel magnetic configuration becomes larger than the current flowing in the antiparallel configuration. As a result, the Rashba coupling
gives rise to a sign change of the TMR, such that its bias-dependent variation ranges from ${\rm TMR}\approx-15\%$ to ${\rm TMR}\approx20\%$ [see
\fig{Fig:TMR_Nonequilibrium}(f)].

We have also analyzed the bias-voltage dependence of the TMR for other values of the dots' levels positions, not satisfying the condition
$\varepsilon=\varepsilon_L=-\varepsilon_R$. As can be seen in \figs{Fig:TMR_Nonequilibrium}(d)--(f), these cases generally exhibit combinations of the effects
discussed above. In particular, for both considered tunings, i.e., $\varepsilon_L=\varepsilon_R=-t_S$ (see the blue curves in \fig{Fig:TMR_Nonequilibrium}) and
$\varepsilon=0$, $\varepsilon_R=-t_S$ (see the green curves in \fig{Fig:TMR_Nonequilibrium}), the TMR oscillates as a function of applied bias voltage and may exhibit
sign changes, with local maxima or minima that appear in the vicinity of the zero bias voltage.

%%%%%%%%%%%%%%%%%%%%%%%%%%%%%%%%%%%
\subsection{Cooper pair splitting efficiency}
\label{Sec:IVC}
%%%%%%%%%%%%%%%%%%%%%%%%%%%%%%%%%%%

\begin{figure}[t]
   \includegraphics[width=1\columnwidth]{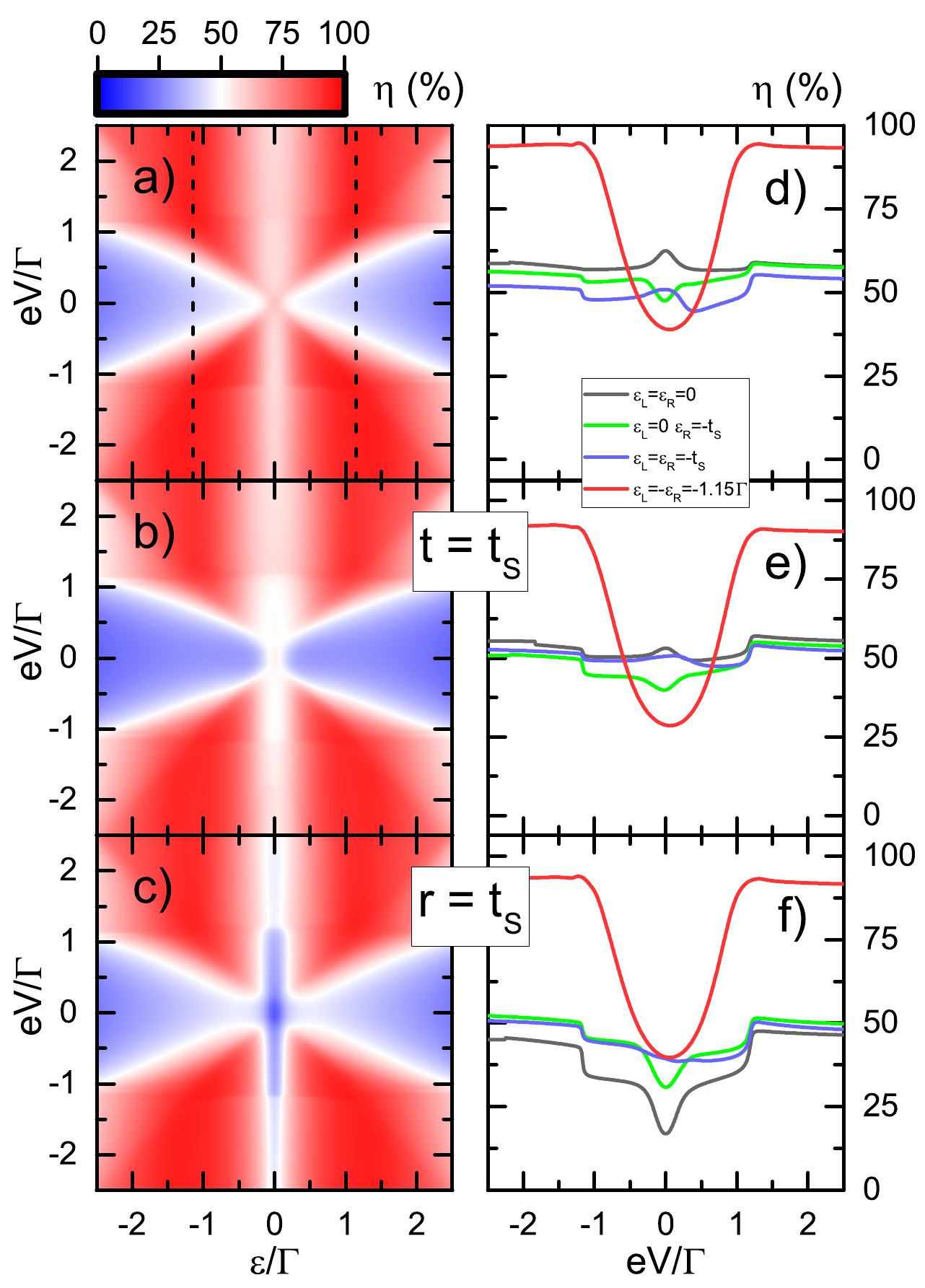}
   \caption{
       The efficiency of the Cooper pair splitting $\eta$
       calculated as a function of the left and right quantum dot
       energy levels $\e=\varepsilon_{L}=-\varepsilon_{R}$
       and the applied bias voltage $V$ (left column).
       The right column shows the bias dependence
       of $\eta$ for indicated values of the dot level positions.
        The first row corresponds to the case without
	 hopping between the dots,
	 while the second (third) row is calculated for
	 $t=t_S$ and $r=0$ ($t=0$ and $r=t_S$).
	This figure is calculated in the
       	case of the antiparallel magnetic configuration
	and for other parameters the same as in \fig{Fig:G2D}.
   }
   \label{Fig:Nonefficiency_equilibrium}
\end{figure}

In \fig{Fig:Nonefficiency_equilibrium} we present the nonequilibrium behavior of the splitting efficiency, which allows for obtaining a deeper insight into different
types of processes responsible for the Andreev transmission through the studied CPS setup. Because the behavior of the splitting efficiency in the parallel
configuration, especially for voltages larger than the inter-dot hopping amplitudes, is qualitatively very similar to that in the antiparallel configuration, here we
only show $\eta$ calculated for the latter situation. As follows from the examination of density plots of $\eta(\varepsilon,V)$ in
\figs{Fig:Nonefficiency_equilibrium}(a)-(c), the splitting efficiency in nonequilibrium situation may acquire very large values for a wide range of bias voltages,
provided that the dots' discrete levels are tuned according to $\varepsilon=\varepsilon_L=-\varepsilon_R$. In particular, when $\varepsilon\approx \pm{\varepsilon_A}$
[see the red cross-sections in \figs{Fig:Nonefficiency_equilibrium}(d)-(f) taken along the dashed line in \fig{Fig:Nonefficiency_equilibrium}(a)], with increasing the
transport voltage the Andreev bound states at $\pm\varepsilon_A$ become gradually activated, leading to an amplification of the nonlocal Andreev reflection processes.
This gives rise to a significant enhancement of the Cooper pair splitting efficiency, which reaches its maximum value $\eta \to 100\%$ above some threshold bias
voltage. The same scenario also holds for finite couplings $t$ and $r$, where it can be seen that once $\varepsilon\approx \pm{\varepsilon_A}$, then the splitting
efficiency is practically insensitive to the variations of the amplitudes $t$ and $r$ (see \fig{Fig:Nonefficiency_equilibrium}). Nevertheless, some slight differences
can still be observed around the zero bias voltage, which demonstrates that the interdot hoppings can increase the number of DAR processes, suppressing thus the
splitting efficiency within the range of a few percents.

Interestingly, when the dot levels are either not detuned at all ($\e=0$)
or detuned such that the condition $\varepsilon_L=-\varepsilon_R$ is not fulfilled any more,
the bias-dependent splitting efficiency becomes drastically suppressed.
This can be clearly seen in \figs{Fig:Nonefficiency_equilibrium}(d)-(f),
where the gray ($\varepsilon=0$),
blue ($\varepsilon_L=\varepsilon_R=-t_S$) and green ($\varepsilon_L=0$, $\varepsilon_R=-t_S$)
lines show that for $|eV|\gtrsim\e_A$,
regardless of the value of the coupling strength $t$ and $r$,
the nonlocal transmission is approximately comparable to the local one,
such that the splitting efficiency oscillates around 50$\%$.
On the other hand, when $|eV|\lesssim\e_A$,
then an interplay between DAR and CAR contributions
in the presence of interdot hopping results in more changes of the splitting efficiency behavior.
The most significant modification of $\eta$
is its Rashba-driven reduction to $\eta\approx$18 $\%$.

%%%%%%%%%%%%%%%%%%%%%%%%%%%%%%%%%%%%%%%%%%%%%%
\section{Summary and conclusions}
\label{Sec:V}
%%%%%%%%%%%%%%%%%%%%%%%%%%%%%%%%%%%%%%%%%%%%%%

In this paper, we have studied the transport properties of a Cooper pair splitter based on triple quantum dots attached to two ferromagnetic contacts and to one
superconducting electrode. In such a setup, the Cooper pairs are extracted by tunneling processes and split into the two arms containing embedded tunable quantum dots
with finite onsite Coulomb correlations, while a large, middle quantum dot is formed in a direct proximity of the superconductor. The system is assumed to work at
sufficiently low temperatures and at voltages smaller than the superconducting energy gap, such that the current is exclusively due to Andreev reflection processes.
Our main focus was on optimizing the parameters of the device to maximize its splitting efficiency, namely, to make the nonlocal crossed Andreev processes through the
both arms of the splitter dominate over the direct Andreev reflections that occur through the same arm of the CPS device. For that, we thoroughly analyzed the effects
of spin-resolved processes, as well as direct and Rashba hopping between the side dots, on the splitting properties of the system, depending on its magnetic
configuration. From the methodological side, we used the Keldysh Green's function approach to study the system's transport properties in both the linear and nonlinear
response regimes.

First of all, we have shown that the Andreev
transport strongly depends on the alignment of magnetic moments
of external ferromagnetic leads and the degree of their spin polarization.
In addition, Andreev reflection processes can be also strongly modified
by the quantum interference between the local and nonlocal processes.
In fact, destructive quantum interference deteriorates
the operation of the CPS by suppressing the transmission of entangled carriers.
A significant manifestation of this effect is observed
in the case of antiparallel magnetic configuration,
especially for electrodes with large spin polarizations.
We have also studied the effects of direct hopping between the side quantum dots
and shown that the quantum interference may be greatly affected by finite hopping amplitude.
All this provides a possibility for controlling and optimizing
the desired properties of the CPS device
by appropriately tuning the interplay between spin polarization of the leads
and the strength of the interdot coupling.

Our main findings encompass among others a nonmonotonic dependence of the splitting efficiency $\eta$ on the spin polarization of the leads, which can be strongly
modified by finite amplitude of hopping between the two side quantum dots. The results revealed a general detrimental impact of the interdot hopping on the splitting
efficiency, with Rashba-type interdot interactions $r$ suppressing $\eta$ more as compared to the direct hopping $t$. However, contrary to this general observation, by
carefully analyzing the splitting efficiency as a function of both $t$ and $r$, we have also identified certain parameter regions where $\eta$ can be greatly enhanced,
reaching $\eta\approx100\%$. In addition, we have considered the dependence of the splitting efficiency in the nonequilibrium transport regime. It turned out that
$\eta$ becomes enhanced in the case when the dots' energy levels are tuned to the position of the Andreev bound states, $\e_L = -\e_R \approx \e_A$, i.e., when the
level of one of the side dots coincides with the Andreev bound state energy $\e_A$, while the other level is aligned at $-\e_A$. In this case, almost perfect splitting
efficiency was found.

Finally, we have also thoroughly analyzed the behavior of the tunnel magnetoresistance,
which is an important quantity in estimating the CAR and DAR contributions
to the total conductance. This is because mainly CAR processes
become affected when the magnetic configuration of the system is varied.
In fact, the sign of the tunnel magnetoresistance as well as
the magnitude of its local maxima and minima
provide an insight into the types of processes that
dominate the total transmission.
We have shown that the magnetoresistive properties
of the Andreev current strongly depend on the parameters
of the device and, especially, on the magnitude and type of hopping
between the side quantum dots. In particular, in equilibrium situations,
direct hopping was shown to enhance the effect of negative TMR.
Moreover, in the nonlinear response regime,
we have found small regions of negative TMR that develop
for voltages larger than the energies of Andreev bound states
and are hardly affected by finite hopping between the dots.

%%%%%%%%%%%%%%%%%%%%%%%%%%%%%%%%%%%%%%%%%%%%%%
\acknowledgments
%%%%%%%%%%%%%%%%%%%%%%%%%%%%%%%%%%%%%%%%%%%%%%

We acknowledge discussions with Piotr Trocha.
This work was supported by the Polish National Science Centre
from funds awarded through the decision No. DEC-2013/10/E/ST3/00213.

%%%%%%%%%%%%%%%%%%%%%%%%%%%%%%%%%%%%%%%%%%%%%%

%\bibliography{Bibliography,BibKB}


\end{document}